\RequirePackage{lineno} 
\documentclass[aps,prc,bibnotes,superscriptaddress,nofootinbib,showpacs,floatfix,singlecolumn]{revtex4}
\usepackage{graphicx} 
\usepackage{float} 
\usepackage{amssymb}
\usepackage{url}
\usepackage[colorlinks,linkcolor=blue,citecolor=blue,filecolor=black,urlcolor=blue]{hyperref}
\usepackage{MnSymbol}
\usepackage{multirow}
\usepackage{bm}
\usepackage{footmisc}

\newcommand{\lr}[1]{\left\langle #1\right\rangle}

\newcommand{\pA}{\mbox{$p$+A}}

\newcommand{\sumet}{\Sigma E_{\textrm T}}

\newcommand{\nchrec}{\mbox{$N_{\mathrm{ch}}^{\mathrm{rec}}$}}
\newcommand{\Ns}{V}
\newcommand{\Nsa}{V_{{a}}}
\newcommand{\Nsb}{V_{{b}}}
\newcommand{\Na}{N_{{a}}}
\newcommand{\Nb}{N_{{b}}}
\newcommand{\na}{n_{{a}}}
\newcommand{\nb}{n_{{b}}}
\newcommand{\bara}[1]{\bar{#1}_{{a}}}
\newcommand{\barb}[1]{\bar{#1}_{{b}}}
\newcommand{\Kn}[2]{K_{#1,{#2}}}
\newcommand{\kn}[2]{k_{#1,{#2}}}
\newcommand{\npart}{N_{\mathrm{part}}}
\newcommand{\npartf}{N_{\mathrm{part}}^{\mathrm{F}}}
\newcommand{\npartb}{N_{\mathrm{part}}^{\mathrm{B}}}

\newcommand{\sqrtnn}{\mbox{$\sqrt{s_{\mathrm{NN}}}$}}

\usepackage{color}
\usepackage{harpoon}
\definecolor{my}{rgb}{1, 0, 0}
\begin{document} 
\title{Centrality fluctuations and decorrelations in heavy-ion collisions}
 \newcommand{\sunysb}{Department of Chemistry, Stony Brook University, Stony Brook, NY 11794, USA}
 \newcommand{\bnl}{Physics Department, Brookhaven National Laboratory, Upton, NY 11796, USA}
\newcommand{\sari}{Shanghai Advanced Research Institute, Chinese Academy of Sciences, Shanghai 201210, China}
\author{Jiangyong Jia}\email[Correspond to\ ]{jiangyong.jia@stonybrook.edu} \affiliation{\sunysb}\affiliation{\bnl}
\author{Chunjian Zhang}\email[Correspond to\ ]{chunjian.zhang@stonybrook.edu}\affiliation{\sunysb}
\author{Jun Xu}\affiliation{\sari}
\date{\today}
\begin{abstract}
The centrality or the number of initial-state sources $V$ of the system produced in heavy ion collision is a concept that is not uniquely defined and subject to significant theoretical and experimental uncertainties. We argue that a more robust connection between the initial-state sources with final-state multiplicity could be established from the event-by-event multiplicity correlation between two subevents separated in pseudorapidity, $N_a$ vs $N_b$. This correlation is sensitive to two main types of centrality fluctuations (CF): 1) particle production for each source $p(n)$ which smears the relation between $V$ and $N_a$ used for experimental centrality, and 2) decorrelations between the sources in the two subevents $V_b$ and $V_a$. The CF is analyzed in terms of cumulants of $V_b$ and $N_b$ as a function of $N_a$, i.e. experimental centrality is defined with $N_a$. We found that the mean values $\lr{V_b}_{N_a}$ and $\lr{N_b}_{N_a}$ increase linearly with $N_a$ in mid-central collisions, but flatten out in ultra-central collisions. Such non-linear behavior is sensitive to the centrality resolution of $N_a$. In the presence of centrality decorrelations, the scaled variances $\langle (\delta V_b)^2\rangle/\lr{V_b}$ and $\langle (\delta N_b)^2\rangle/\lr{N_b}$ are found to decrease linearly with $N_a$ in mid-central collisions, while the $p(n)$ leads to another sharp decrease in the  ultra-central region. The higher-order cumulants of $V_b$ and $N_b$ show interesting but rather complex behaviors which deserve further studies. Our results suggest that one can use the cumulants of the two-dimensional multiplicity correlation, especially the mean and variance, to constrain the particle production mechanism as well as the longitudinal fluctuations of the initial-state sources.
\end{abstract}
\pacs{25.75.Dw} 
\maketitle
\section{Introduction}\label{sec:1}
In heavy-ion collisions, centrality is an important concept for characterizing the size of the produced fireball. The centrality of an event is often represented by the number of particle production sources $\Ns$ (for example participating nucleons or constituent quarks) in the initial state. These sources are then used to define other important geometry quantities, such as the nuclear overlap function~\cite{Eskola:1988yh,dEnterria:2003xac} for jet quenching studies~\cite{Qin:2015srf,Connors:2017ptx} and eccentricities~\cite{Teaney:2010vd,Alver:2008zza} for collective flow studies.~\cite{Luzum:2013yya,Jia:2014ysa} Since the sources in an event are not directly measurable, a Glauber model is used to calculate $\Ns$ and relate it to the final observed particle multiplicity $N$, with $N=\sum_{i=1}^{\Ns} n_i$ calculated as a sum of the multiplicity for each source $n_i$ sampled independently from a common distribution $p(n)$~\cite{Abelev:2013qoq,Miller:2007ri,Adler:2013aqf}. Due to fluctuations in the particle production, events selected with the same $\Ns$ can have different values of $N$ and vice versa. The fluctuation of sources at fixed multiplicity, often referred to as the ``volume fluctuations'', is an irreducible ``centrality fluctuations'' (CF) or centrality resolution~\cite{Jeon:2003gk,Skokov:2012ds,Luo:2013bmi,Zhou:2018fxx}. The main issue concerning centrality is to understand how the final-state $p(N)$ is related to the source distribution $p(\Ns)$ and particle production for each source $p(n)$. 

\begin{figure}[h!]
\begin{center}
\includegraphics[width=1\linewidth]{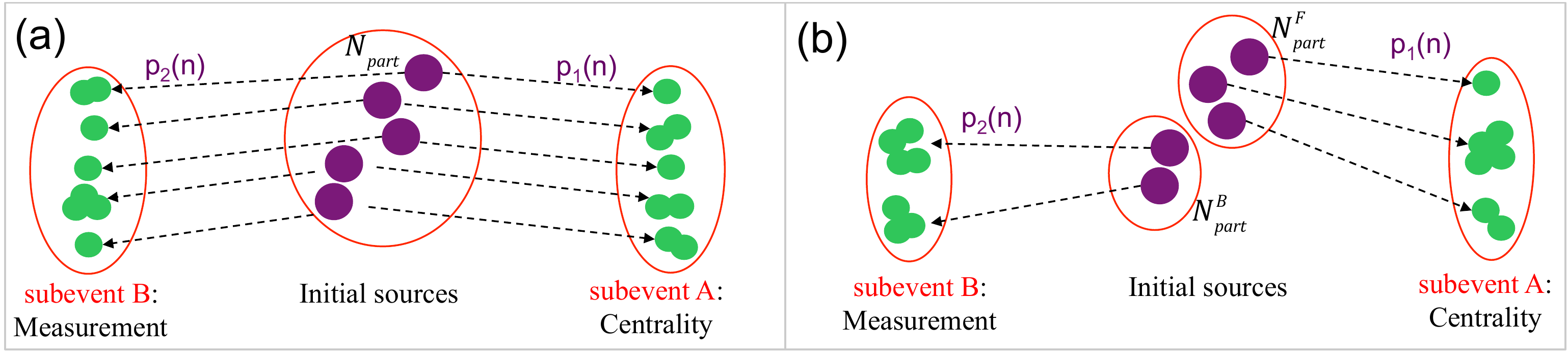}
\end{center}
\vspace*{-0.5cm}\caption{\label{fig:1} Schematic illustration of the relation between sources (large circles) and final-state particles (small circles) produced in subevent A for centrality determination and subevent B for measurements for two cases: 1) same sources, wounded nucleons $\npart$, for both subevents (panel a), and 2) different sources, forward-going and backward-going wound nucleons $\npartf$ and $\npartb$, for the two subevents (panel b). The second scenario is particularly relavent for fixed-target experiments, which often use the forward multiplicity for centrality and mid- or backward rapidity for multiplicity measurements.}\vspace*{-0.3cm}
\end{figure}

The centrality or volume of an event is often assumed to be a global concept, long-range in pseudorapidity ($\eta$): a central event should be a central event independent of $\eta$. This assumption can be checked by studying the the correlation of multiplicities between two different $\eta$ windows, also referred to as forward-backward (FB) multiplicity correlation~\cite{Back:2006id,Abelev:2009ag,Bialas:2010zb,Bzdak:2012tp,Jia:2015jga}. In experimental analysis, the observed centrality is usually defined as the particle multiplicity $\Na$ in a forward pseudorapidity window A, and the measurement is performed using particles with total multiplicity $\Nb$ in a different pseudorapidity window B, usually around mid-rapidity. If the system is boost invariant and same sources are responsible for particles in both subevents (see Fig~\ref{fig:1}(a)), the particle multiplicity distribution in subevent B for fixed $\Na$ can be expressed as,
\begin{align}\label{eq:1}
p(\Nb)_{\Na}= \sum_{\Ns} p(\Nb)_{\Ns}p(\Ns)_{\Na}.
\end{align}
The distribution $p(\Ns)_{\Na}$, reflecting the extent of the CF, is controlled by the smearing from $p(n)$ in subevent A. Narrow $p(\Ns)_{\Na}$ distribution would imply poor centrality resolution in subevent A and vice versa. Eq.~\eqref{eq:1} shows that the CF arising from subevent A directly influences the multiplicity fluctuations in subevent B, and this is the picture assumed by most previous analyses~\cite{Miller:2007ri,Abelev:2013qoq,Adler:2013aqf}. 

Recently, several studies also consider the possibility that the number of sources and their transverse distributions may fluctuate along $\eta$ in a single event. This is because the number of forward-going and backward-going participating nucleons, i.e. $\npartf$ and $\npartb$, are not the same in a given event~\cite{Jia:2015jga,Jia:2014ysa}, and they contribute differently in the forward (backward) rapidity.  This forward-backward asymmetry is realized in one of the two ways in dynamical models: 1) In string picture~\cite{Bozek:2015bna,Pang:2015zrq,Shen:2017bsr}, each string originates from a participant nucleon and extends to the opposite direction. Therefore, the number of strings, their lengths and end-points fluctuate in rapidity. 2) In a gluon saturation picture, the sub-nucleonic degree-of-freedoms evolve with rapidity~\cite{Schenke:2016ksl}. In the forward rapidity, the projectile nucleons are dominated by a few large-$x$ partons, while the target nucleons are expected to contribute mainly low-$x$ partons. In this case, the Eq.~\eqref{eq:1} can be revised as,
\begin{align}\label{eq:2}
p(\Nb)_{\Na}= \sum_{\Nsa,\Nsb} p(\Nb)_{\Nsb}p(\Nsb)_{\Nsa}p(\Nsa)_{\Na}.
\end{align}
The fluctuation between the number of sources in the two subevents, denoted by $p(\Nsb)_{\Nsa}$, can weaken the centrality correlation between different rapidities (see Fig~\ref{fig:1}(b)). Such ``centrality decorrelations'' effects is expected to have a strong influence on the correlation of multiplicities between two different $\eta$ windows.

Eqs.~\eqref{eq:1} and \eqref{eq:2} show that information about centrality fluctuations and particle productions is encoded in a set of one-dimensional distributions, which are customarily analyzed via multiplicity cumulants of $p(x)$ ($x=N$, $n$ or $\Ns$): mean $\bar{x}\equiv\lr{x}$, variance $\lr{(x-\bar{x})^2}$,  skewness $\lr{(x-\bar{x})^3}$,... In this paper, we use the reduced form to scale out the overall system size:
\begin{align}\label{eq:3}
k_1\{p(x)\}=\bar{x}, k_2\{p(x)\}=\lr{(x-\bar{x})^2}/\bar{x}, k_3\{p(x)\}=\lr{(x-\bar{x})^3}/\bar{x},....
\end{align}
Previous studies mainly focused on the scaling behavior of the mean multiplicity per source, $N/\lr{\Ns}$~\cite{Adler:2013aqf,Lacey:2016hqy,Loizides:2016djv,Bozek:2016kpf,Acharya:2018hhy,Rohrmoser:2019xis}. Our paper extends these studies to higher-order moments/cumulants of $p(N)$ and $p(\Ns)$, and their dependences on particle production $p(n)$. These cumulants reveal detailed structures of the two-dimensional correlation between $\Nb$ and $\Na$, and provide constraints on $p(n)$, $p(\Nsa)$ and $p(\Nsb)_{\Nsa}$. 

A good understanding of CF is also important for other areas of heavy-ion physics: for any physics observable $x$ that varies with $\Ns$, its fluctuation $p(x)$ should be sensitive to $p(\Ns)$~\cite{Zhou:2018fxx}. The CF is one of the main background sources in the ongoing search for the critical end point in the QCD phase diagram based on the fluctuations of conserved quantities~\cite{Luo:2017faz,Li:2017via}. The CF also strongly affects the fluctuations of harmonic flow~\cite{Zhou:2018fxx}, and is responsible for the observed sign-change behavior in ultra-central A+A collisions (UCC)~\cite{Aaboud:2019sma}. Understanding the longitudinal fluctuations of sources and the resulting centrality decorrelation effects can also help to describe previous measurements of harmonic flow decorrelations in $\eta$~\cite{Pang:2015zrq,Khachatryan:2015oea,Aaboud:2017tql,Aad:2020gfz}.

In this paper, we study the influence of centrality smearing from particle production (Fig~\ref{fig:1}a)) and centrality decorrelations from the initial sources (Fig~\ref{fig:1}b)) on the event-by-event forward-backward multiplicity fluctuations. The study is based on an independent source Glauber model with particle production tuned to reproduce the ATLAS data~\cite{Aaboud:2019sma}, and the initial number of sources is chosen to be the $\Ns=\npart$ for the scenario shown in Fig~\ref{fig:1}a) and $\Nsa\equiv\npartf$ and $\Nsb\equiv\npartb$ for the scenario shown in Fig~\ref{fig:1}b). We found that the fluctuations in the UCC region are very sensitive to the source distribution $p(\Ns)$ and particle production $p(n)$, due to the steeply falling distributions (upper boundary effect)~\cite{Skokov:2012ds,Xu:2016qzd,Bzdak:2016jxo}. We also found that the multiplicity dependence of the mean and variance of $p(\Ns)$ have a very simple interpretation in terms of the two types of centrality fluctuations, and they can constrain the longitudinal structure of initial sources and particle production mechanism via comparison between data and model. The structure of the paper is the following. Section~\ref{sec:2} and \ref{sec:3} discuss the expected general behavior of the CF in the presence of multiplicity smearing and centrality decorrelations, respectively. After a brief description of the Glauber model setup in Section~\ref{sec:4}, the results for the two types of CF are given in Section~\ref{sec:5} and \ref{sec:6}, respectively. The summary of the results are given in Section~\ref{sec:7}. The Appendix~\ref{sec:app1} gives a derivation of the formulas used in this paper, and some additional results are given in Appendix~\ref{sec:app2}.

\section{centrality fluctuations due to multiplicity smearing}\label{sec:2}
The CF arising from multiplicity smearing effects (Fig.~\ref{fig:1}a)) is considered in an independent source model framework, by assuming that the total particle multiplicity is a sum of particles from each source $N \equiv \sum_{i=1}^{\Ns} n_i$ sampled from $p(n)$. In this case, one can show that cumulants of $p(\Nb)_{\Ns}$ in Eq.~\eqref{eq:1} is related to $p(n)$ by a factor $\Ns$~\cite{Skokov:2012ds}, i.e. $\lr{\Nb}_{\Ns} = \Ns \lr{n}\;, \lr{\delta \Nb^2}_{\Ns} = \Ns \lr{\delta n^2}\;, \lr{\delta \Nb^3}_{\Ns} = \Ns \lr{\delta n^3}\;,....$, and that
\begin{align}\label{eq:4}
k_m\{p(\Nb)_{\Ns}\} = k_m\{p(n)\}, m=2,3,4.....
\end{align}
From this, Skokov et al.~\cite{Skokov:2012ds} find the following relations connecting the cumulants for the three distributions in Eq.~\eqref{eq:1}:
\begin{align}\nonumber
\bar{\Nb} &= \barb{n}\lr{\Ns}_{\Na}\;, ( \bar{\Nb} \equiv  \Kn{1}{b}\;, \barb{n} \equiv  \kn{1}{b} \;, \lr{\Ns}_{\Na} \equiv  \kn{1}{\Ns|\Na})\\\nonumber
K_{2,\mathrm{b}} &= k_{2,\mathrm{b}}+\barb{n}k_{2,\Ns|\Na}\;,\\\nonumber
K_{3,\mathrm{b}} &= k_{3,\mathrm{b}}+3k_{2,\mathrm{b}}\barb{n}k_{2,\Ns|\Na}+\barb{n}^2k_{3,\Ns|\Na},\\\label{eq:5}
K_{4,\mathrm{b}} &= k_{4,\mathrm{b}}+(4k_{3,\mathrm{b}}+3k_{2,\mathrm{b}}^2)\barb{n}k_{2,\Ns|\Na}+6k_{2,\mathrm{b}}\barb{n}^2k_{3,\Ns|\Na}+\barb{n}^3k_{4,\Ns|\Na},
\end{align}
where we have used a simpler notation, $k_n\{p(N)\}\rightarrow K_n$, $k_n\{p(N)_{\Ns}\}\rightarrow k_n$ and $k_n\{p(\Ns)\}\rightarrow k_{n,\Ns}$.

\begin{figure}[h!]
\begin{center}
\includegraphics[width=0.7\linewidth]{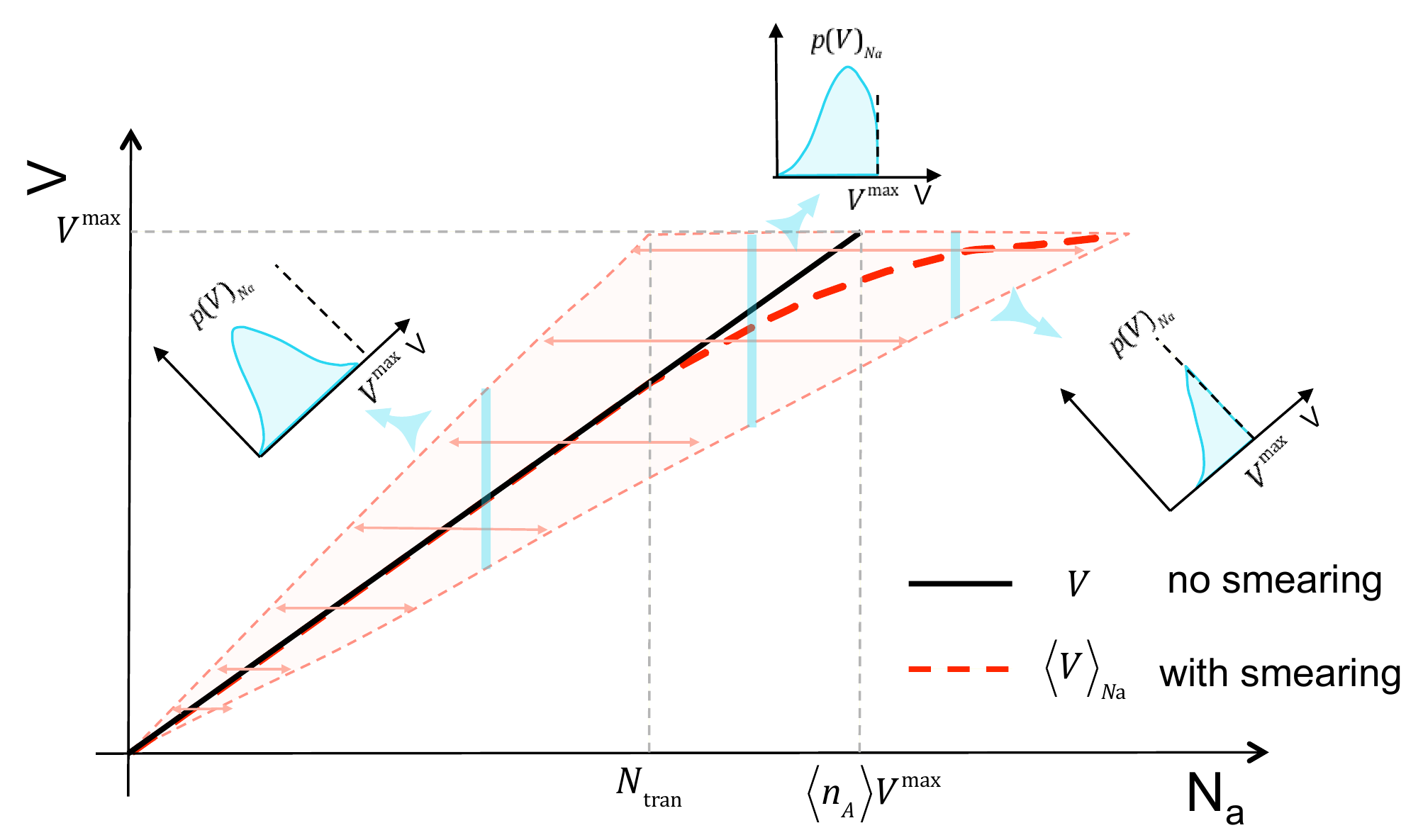}
\end{center}
\caption{\label{fig:2} Schematic illustration of correlation between $\Ns$ and $\Na$ before (solid line) and after multiplicity smearing (shaded average) which broadens the $\Na$ distribution at fixed $\Ns$. The average $\Ns$ for fixed $\Na$, $\lr{\Ns}_{\Na}$ (red dashed line), deviates from a linear relation in ultra-central collisions. The insert distributions show the $\Ns$ distribution for fixed $\Na$ in mid-central collisions (left panel), central collisions (middle panel) and ultra-central collisions (right panel).}
\end{figure}

The behavior of centrality cumulants $k_{n,\Ns|\Na}$ for $p(\Ns)_{\Na}$ in Eq.~\eqref{eq:1} can be understood qualitatively from the correlation between $\Ns$ and $\Na$ in Fig.~\ref{fig:2}. In the absence of smearing from particle production, i.e. by setting $p(\na)=\delta(\na-\bara{n})$, $\Ns$ and $\Na$ are related by a linear function $\Na=\Ns \bara{n}$ (thick solid line). In the presence of smearing, $p(\na)$ broadens the correlation along the $x$-axis (shaded area). As a result, the $\lr{\Ns}_{\Na}$ deviates from the linear relation in the UCC region (thick dashed line). This deviation arises because the distribution $p(\Ns)_{\Na}$ is bounded above by $\Ns^{\mathrm{max}}$ (twice of the atomic number for $\npart$), leading to an asymmetric shape in the UCC region as illustrated by the middle and right insert panels. For even larger $\Na$ values, the $p(\Ns)_{\Na}$ becomes narrower and more asymmetric, and eventually converges to $\Ns^{\mathrm{max}}$. For this reason, the cumulants of $p(\Ns)$ as a function of $\Na$ are strongly modified in the UCC collisions: in addition to the departure of the mean from the linear relation, the variance of $p(\Ns)_{\Na}$ is expected to approach zero, the skewness of $p(\Ns)_{\Na}$ becomes negative and then approaches zero, and the kurtosis of $p(\Ns)_{\Na}$ oscillates from negative to positive then approaches zero. Outside the UCC region, where $p(\Ns)_{\Na}$ is not constrained by the upper boundary effect (left insert panel), the $\lr{\Ns}_{\Na}$ is expected to overlap with the solid line, and the higher-order cumulants of $p(\Ns)_{\Na}$ are determined mainly by $p(\na)$. In this paper, we focus on the first four cumulants of CF, as their importance can be clearly identified from the shape of $p(\Ns)_{\Na}$.

The origin of CF and the behavior of centrality cumulants $k_{n,\Ns|\Na}$ for $n=2$--4 were studied in detail in Ref.~\cite{Zhou:2018fxx}. In the UCC region, the moments of CF (and therefore $k_{n,\Ns|\Na}$) are driven mainly by the shape of $p(\Ns)$,
\begin{eqnarray}
\label{eq:6}
\lr{(\delta\Ns)^k}_{\Na}\approx \int (\delta\Ns)^k\frac{1}{\sqrt{2\pi\hat{\sigma}^2\Ns}}  \mathrm{exp}({-\frac{(\Ns-\lr{\Ns}_{\scriptscriptstyle{\Na}})^2}{2\hat{\sigma}^2\Ns}}) p(\Ns) d\Ns\;.
\end{eqnarray}
where we assume $p(\Na)_{\Ns} \approx \frac{1}{\sqrt{2\pi \hat{\sigma}^2 \Ns}} e^{-\left.(\Na-\bara{n}\Ns)^2\right/2\hat{\sigma}^2\Ns}$ via the central-limit theorem for large $\Ns$, and $\hat{\sigma}^2=\lr{(\delta \na)^2}/\bara{n}^2$ is the relative width for $p(\na)$. In the mid-central region where $p(\Ns)$ is slowly changing, $k_{n,\Ns|\Na}$ are found to be approximately constant, and the scaled variance of the multiplicity fluctuation in subevent B is sensitive to both $p(\na)$ and $p(\nb)$, 
\begin{align}
\label{eq:7}
\frac{\lr{(\delta\Nb)}^2}{\barb{N}^2} = \frac{1}{\lr{\Ns}_{\Na}}\left(\frac{\lr{(\delta\nb)^2}}{\barb{n}^2}+\frac{\lr{(\delta\na)^2}}{\bara{n}^2}\right).
\end{align}
This relation implies that for mid-central collisions and within the independence source picture, the variance of the measured multiplicity fluctuations has the same functional dependence on $p(\na)$ and $p(\nb)$ in the two subevents.

\section{Centrality decorrelations due to longitudinal fluctuations}\label{sec:3}
Next we consider the CF due to centrality decorrelations as illustrated by Fig.~\ref{fig:1}b). Equation~\eqref{eq:2} can be decomposed into two equations
\begin{align}\label{eq:2a}
p(\Nb)_{\Na}&= \sum_{\Nsb} p(\Nb)_{\Nsb}p(\Nsb)_{\Na}\\\label{eq:2b}
p(\Nsb)_{\Na}&= \sum_{\Nsa} p(\Nsb)_{\Nsa}p(\Nsa)_{\Na}.
\end{align}
The cumulants for the first equation are given by Eq.~\eqref{eq:5}. The second equation contains the effects of centrality decorrelations from $p(\Nsb)_{\Nsa}$, which, unlike the $p(\Nb)_{\Nsb}$, is not described by the independent source picture (i.e $\Nsb$ is not an independent sum of $\Nsa$ number of sources via some common distribution). Indeed, we find that the cumulants of $p(\Nsb)_{\Nsa}$ are not strictly proportional to $\Nsa$. However, by assuming the departure from such proportionality is small, we have derived the relations including the first-order correction (see Appendix~\ref{sec:app1}):
\begin{align}\nonumber
&\bar{V}_{b} = \kn{1}{ab} \lr{\Nsa}_{\Na}\;, \\\nonumber
&k_{2,\Nsb} = k_{2,ab}+\frac{k_{1,ab}'^2}{k_{1,ab}} k_{2,\Nsa|\Na}\;, \\\nonumber
&k_{3,\Nsb} = k_{3,ab}+3k_{2,ab}'k_{1,ab}'k_{2,\Nsa|\Na}+\frac{k_{1,ab}'^3}{k_{1,ab}}k_{3,\Nsa|\Na}\;,\\\label{eq:8}
&k_{4,\Nsb} = k_{4,ab}+(4k_{3,ab}'k_{1,ab}'+3k_{2,ab}'^2k_{1,ab})k_{2,\Nsa|\Na}+6k_{2,ab}'k_{1,ab}'^2k_{3,\Nsa|\Na}+\frac{k_{1,ab}'^4}{k_{1,ab}}k_{4,\Nsa|\Na}\;.
\end{align}
The $k_{n,ab}$ are the reduced cumulants for $p(\Nsb)_{\Nsa}$, while $k_{n,ab}'$ also include the leading-order correction,
\begin{align}\nonumber
&k_{1,ab}' = \frac{\partial (k_{1,ab} \Nsa)}{\partial \Nsa } =  k_{1,ab}+\Nsa \frac{\partial k_{1,ab}}{\partial \Nsa }\;,\\\label{eq:9}
&k_{n,ab}' = \frac{\partial (k_{1,ab}k_{n,ab} \Nsa)}{k_{1,ab}\partial \Nsa } =  k_{n,ab}+\Nsa \frac{\partial (k_{1,ab}k_{n,ab})}{k_{1,ab} \partial \Nsa }\;, n\geq 2
\end{align}

The total multiplicity cumulants in subevent B, $\Kn{n}{b}$, can be obtained by replacing the $k_{n,\Ns|\Na}$ in Eq.~\eqref{eq:5} by Eq.~\eqref{eq:8}. If the centrality resolution of subevent A is very good, $k_{n,\Nsa|\Na}\approx 0$ and $k_{n,\Nsb} \approx k_{n,ab}$ for $n>1$, and we recovers the Eq.~\eqref{eq:5}. In this case, the behavior of $\Kn{n}{b}$ is dictated by the FB fluctuations.

\section{Glauber model setup}\label{sec:4}
For quantitative study of the effects of multiplicity smearing and the longitudinal centrality decorrelations, we follow the implementation of our previous work~\cite{Zhou:2018fxx}, which is described briefly here. The number of sources in subevent A and B are chosen to be $\npartf$ and $\npartb$ calculated for each event in a standard Glauber model framework~\cite{Miller:2007ri}. The nucleons are assumed to have a hard-core of 0.3 fm in radii, their transverse positions are generated according to the Woods-Saxon distribution as provided by Ref.~\cite{Loizides:2014vua}. A nucleon-nucleon cross-section of $\sigma=68$~mb is used to simulate the collisions at $\sqrtnn=5.02$ TeV. Since the $p(\Ns)$ distribution as well as the correlation between $\npartf$ and $\npartb$ change with system size, we studied several nuclei covering a broad range of the $\Ns$: Xe+Xe, Cu+Cu, S+S and O+O collisions, with total number of nucleons $2A=$ 258, 126, 64, 32 respectively. 

The particle productions for each source are chosen to follow the negative binomial distribution (NBD):
\begin{eqnarray}
\label{eq:10}
p(n) = \frac{(n+m-1)!}{(m-1)!n!} p^n(1-p)^m,\; p = \frac{\bar{n}}{\bar{n}+m}
\end{eqnarray}
where $\bar{n}$ is the average number of particles in a given subevent. One important quantity is the relative width $\hat{\sigma}$: $\hat{\sigma}^2 \equiv \left.\lr{(\delta n)^2}\right/\bar{n}^2$, which controls the strength of fluctuations for each source.

Table~\ref{tab:1} lists the three NBD parameter sets used to produce particles in subevents A and B, taken directly from Ref.~\cite{Zhou:2018fxx}. The Par0 and Par1 are adjusted to approximately describe the shapes of the experimental distributions of $\nchrec$ ($|\eta|<2.5$) and $\sumet$ ($3.2<|\eta|<4.9$) from the ATLAS Collaboration~\cite{Aaboud:2019sma}, while the Par2 corresponds to a case with much larger fluctuations. 
\begin{table}[!h]
\centering
\caption{The three parameters sets for the NBD (Eq.~\eqref{eq:10}) used for modeling the particle production in the wounded nucleon model from Ref.~\cite{Zhou:2018fxx}. The corresponding values of reduced cumulants $k_2$, $k_3$ and $k_4$ are also listed.}
\label{tab:1}
\begin{tabular}{|c|c|c|c|c|c|c|c|}\hline 
         &   $p$ &   $m$ & mean $k_1=\bar{n}$& RMS/mean $\hat{\sigma}$ & $k_2$ & $k_3$ & $k_4$ \\\hline
Par0     & 0.688 & 3.45  &      7.6     & 0.65 & 3.2   & 17.3 & 139\\\hline
Par1     & 0.831 & 1.55  &      7.6     & 0.88 & 5.9   & 63.8 & 1031\\\hline
Par2     & 0.928 & 0.593 &      7.6     & 1.35 & 13.8  & 368  & 14692\\\hline
\end{tabular}
\end{table}

The calculation of cumulants follows the standard procedure. Each A+A event has two subevents: subevent A for centrality selection and subevent B for the calculation of multiplicity cumulants. The particle multiplicities in these two subevents, $\Na$ and $\Nb$, are generated independently from 1) the same sources $\Nsa=\Nsb=\npart$ for the study of multiplicity smearing effect or 2) $\Nsa=\npartf$ and $\Nsb=\npartb$ for the study of longitudinal centrality decorrelations. The generated events are divided into classes according to $\Na$. The cumulants are first calculated from $\Ns$ and $\Nb$ distributions for events with the same $\Na$, which are then combined into broader $\Na$ ranges to reduce the statistical uncertainty. In the following, we first discuss the behavior of cumulants by assuming $\Nsa=\Nsb=\npart$, we then show results for cumulants calculated by assuming $\Nsa=\npartf$ and $\Nsb=\npartb$.

\section{Results on centrality fluctuations from particle production}\label{sec:5}
\begin{figure}[h!]
\begin{center}
\includegraphics[width=0.72\linewidth]{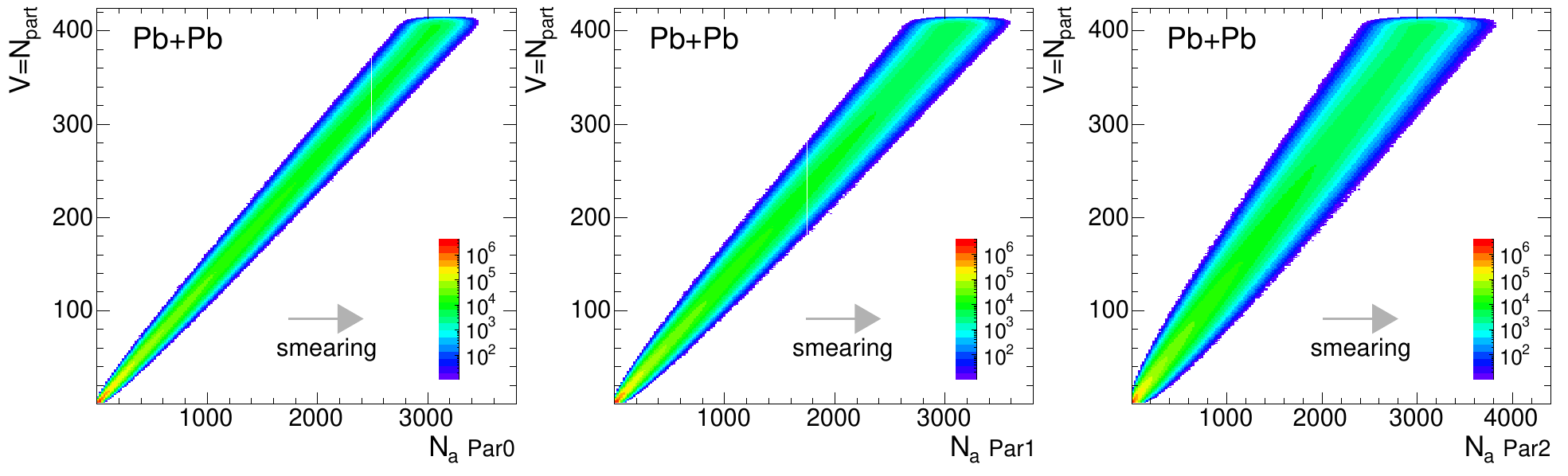}\\
\vspace*{-0.2cm}\includegraphics[width=1\linewidth]{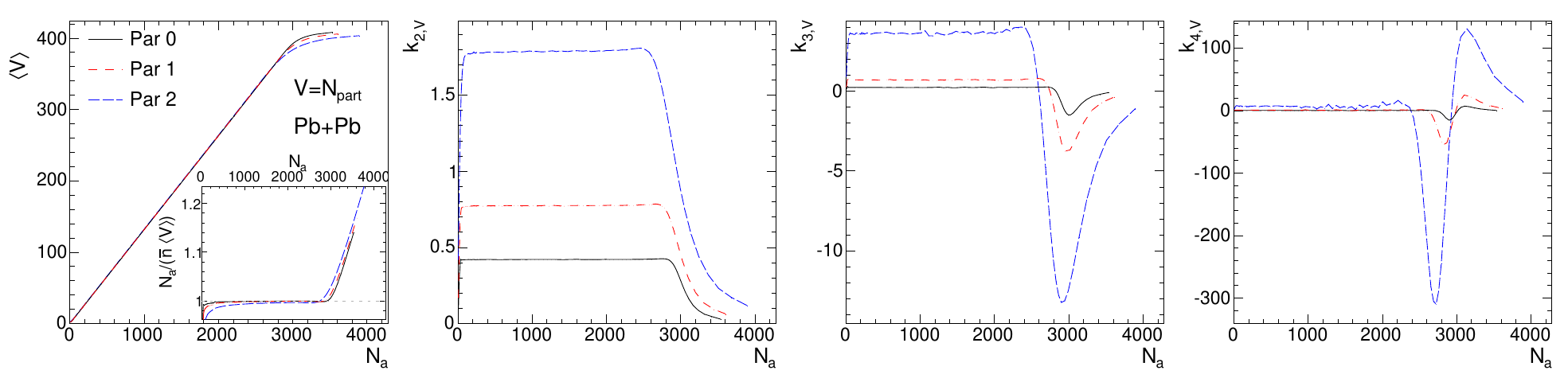}
\end{center}
\vspace*{-0.4cm}\caption{\label{fig:3} Top row: the correlation between $\Ns\equiv\npart$ and $\Na$ in Pb+Pb collisions generated with Par0 (left), Par1 (middle) and Par2 (right) parameter sets from Table~\ref{tab:1}. Bottom row: the corresponding cumulants of $\Ns$, $\kn{n}{\Ns}$, calculated for the three parameter sets, with  $n=1$, 2, 3, and 4 from left to right panels. The insert panel shows the ratio of $\Na/(\bar{n}\lr{V})$, to quantify the deviation of particle per source from $\bar{n}$ (or from the linear relation).}
\end{figure}

We first consider the case when participant nucleons are used as common sources for subevents A and B, $\Nsa=\Nsb=\npart\equiv\Ns$. Figure~\ref{fig:3} illustrates the behavior of the centrality fluctuations by selecting events with fixed $\Na$ in Pb+Pb collisions. The top three panels show the correlation between $\Ns$ and $\Na$ generated with Par0, Par1 and Par2 in Table~\ref{tab:1}. The centrality cumulants $k_{n,\Ns}$ for distribution $p(\Ns)_{\Na}$ are calculated and shown in the bottom panels.

The behavior of these cumulants follows the naive expectation of Fig.~\ref{fig:2}. The $k_{1,\Ns}=\lr{\Ns}$ is proportional to $\Na$, except in UCC region where it turns over. For larger $\hat{\sigma}$ value, this turn-over starts earlier and extends to larger $\Na$ range, as expected from a poorer centrality resolution. The insert panel shows the ratio $\Na/(\bar{n}\lr{V})$, to quantify the deviation of particle per source from $\bar{n}$. This ratio should be unity in the absence of centrality smearing effects. The ratio exhibits a suppression in the peripheral region and an enhancement in the central region, as expected from the boundary effects. We notice that in the central region, the ratios for different parameter sets are nearly parallel to each other. We found that the slopes of the ratios are mainly controlled by the shape of the $p(\npart)$ distributions for the UCC events. The larger $\hat{\sigma}$ value only shifts the turn-over point to a smaller $\Na$ value and extends the suppression region to a larger $\Na$ value, but has little impact on the slope.

The behavior of higher-order centrality cumulants follows what was observed before in Ref.~\cite{Zhou:2018fxx}.  The $k_{n,\Ns}$ are flat in mid-central collisions but show strong variations in central region: the $k_{2,\Ns}$ decreases to 0,  the $k_{3,\Ns}$ decreases to negative minimum values and then approaches 0 from below, while the $k_{4,\Ns}$ decreases to a negative minimum, increases to reach a positive maximum and then decreases to approach 0. The larger smearing associated with Par2 further enhances the magnitudes of $k_{n,\Ns}$ and their oscillating behaviors in the ultra-central region.

\begin{figure}[h!]
\begin{center}
\vspace*{-0.4cm}\includegraphics[width=0.72\linewidth]{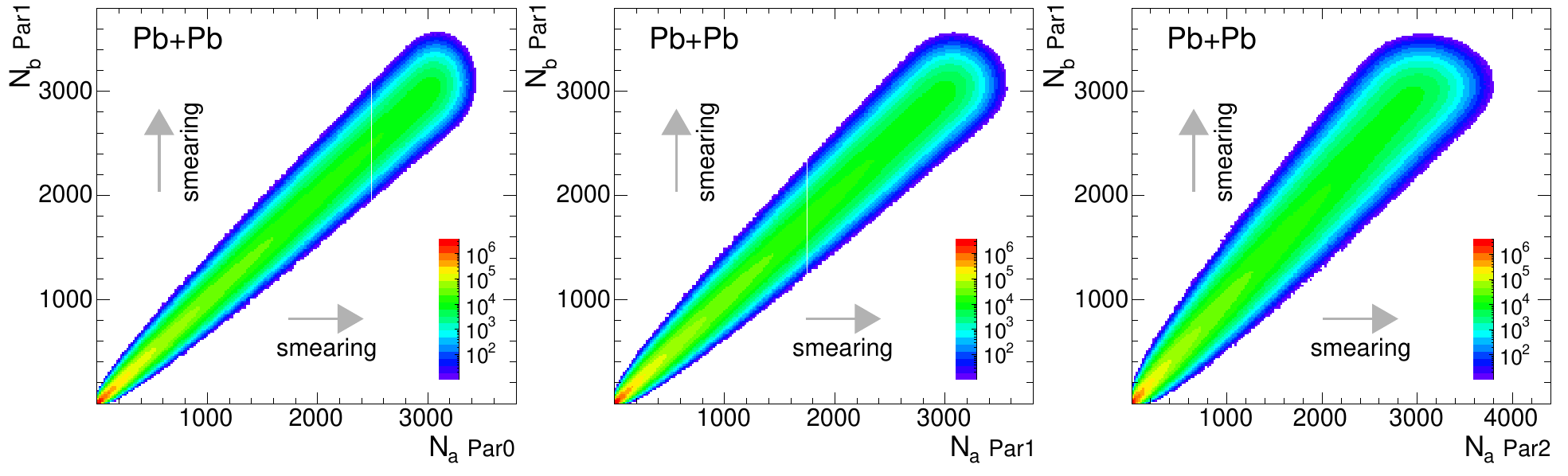}\\
\includegraphics[width=1\linewidth]{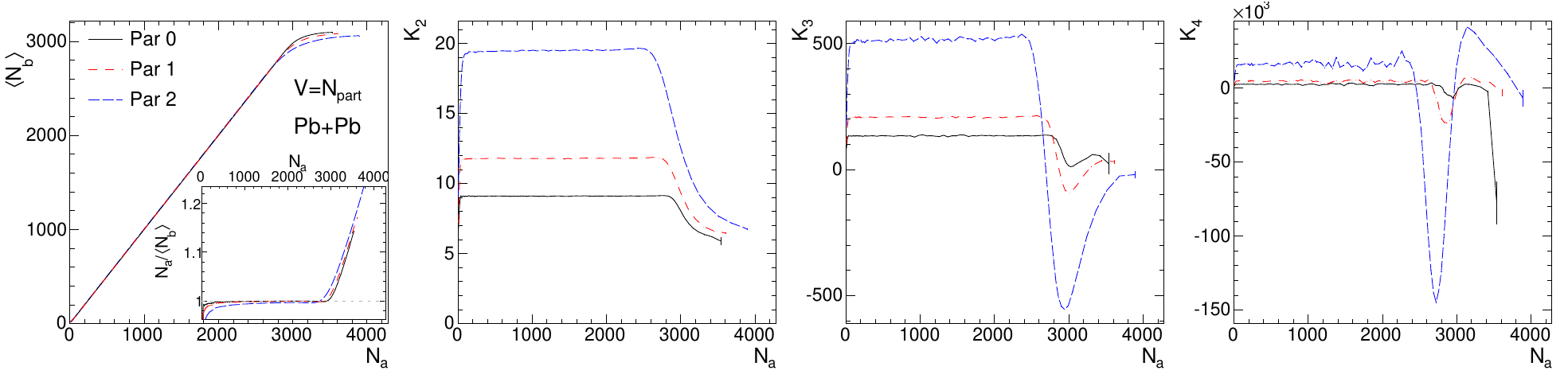}
\end{center}
\caption{\label{fig:4} Top row: the correlation between $\Nb$ generated with Par1 and $\Na$ generated with Par0 (left), Par1 (middle) and Par2 (right) in Pb+Pb collisions. Bottom row: the corresponding cumulants of $\Nb$, i.e. $\Kn{n}{\Nb}$, calculated from the three distributions in the top row, with $n=1$, 2, 3, and 4 from left to right panels. The insert panel show the ratio of $\Na/\lr{\Nb}$, to quantify the deviation of $\Na$ dependence of $\lr{\Nb}$ from the linear relation.}
\end{figure}

The source distributions for events with fixed $\Na$, $p(\Ns)_{\Na}$, is then used to produce the distribution of $\Nb$ according to a common distribution $p(\nb)$ for each source. The top panels of Fig.~\ref{fig:4} shows the correlation between $\Nb$ and $\Na$, obtained by smearing the 2D distributions in Fig.~\ref{fig:3} along the $y$-axis using the Par1 for $p(\nb)$. The cumulants of $p(\Nb)_{\Na}$, $\Kn{n}{b}$, are calculated and shown in the bottom panels. According to Eq.~\eqref{eq:5}, they are expressed as a linear combination of $k_{n,\Ns}$,
\begin{align}\nonumber
\bar{\Nb} &= 7.6\lr{\Ns}\;\\\nonumber
K_{2,\mathrm{b}} &= 5.9+7.6\kn{2}{\Ns}\;,\\\nonumber
K_{3,\mathrm{b}} &= 63.8 +135 \kn{2}{\Ns}+57.8 \kn{3}{\Ns},\\\label{eq:11}
K_{4,\mathrm{b}} &= 1031+2733 \kn{2}{\Ns}+2045\kn{3}{\Ns}+439\kn{4}{\Ns},
\end{align}
where the coefficients are determined from the $k_{n}$ for Par1 in Table~\ref{tab:1}. Eq.~\eqref{eq:11} shows that the $\lr{\Nb}$ is related to $\lr{\Ns}$ in Fig.~\ref{fig:3} by a constant scale factor. The $\Kn{2}{b}$ is related to $k_{2,\Ns}$ by a constant offset and a constant scale factor, etc. In general, one could isolate the $\kn{n}{\Ns}$ iteratively order-by-order: $\kn{2}{\Ns}$ can be extracted from $K_{2,\mathrm{b}}$ by identifying and subtracting a constant, which then can be subtracted from $K_{3,\mathrm{b}}$ to isolate the $\kn{3}{\Ns}$, etc. Note that this is only true in the independent source model framework and considering only multiplicity smearing effects.

\begin{figure}[h!]
\begin{center}
\includegraphics[width=1\linewidth]{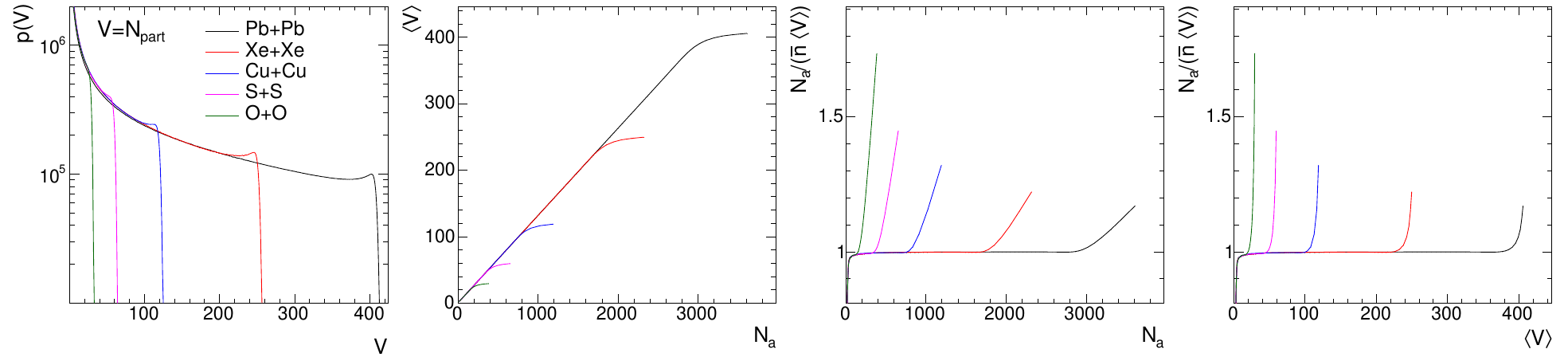}\\
\includegraphics[width=1\linewidth]{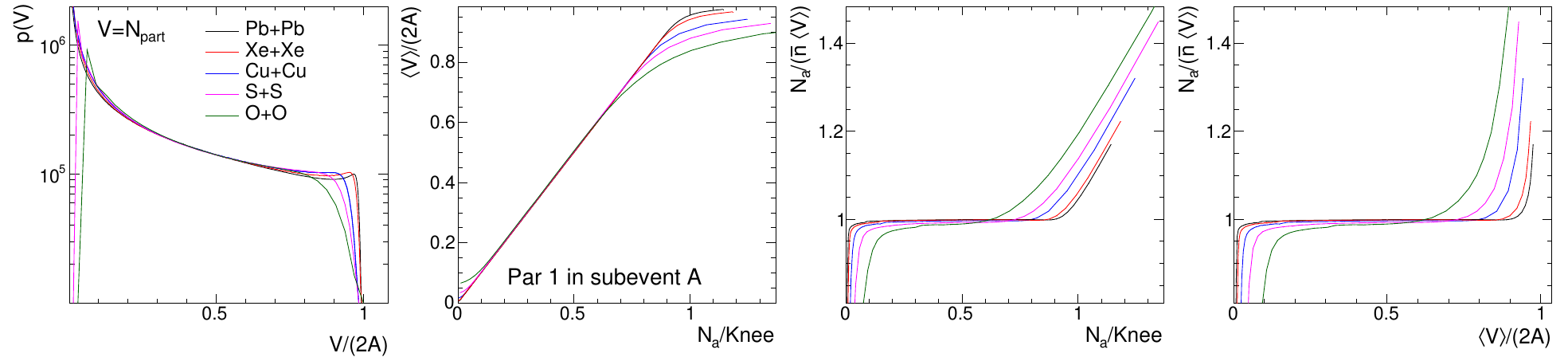}
\end{center}
\vspace*{-0.4cm}\caption{\label{fig:5} The distribution of sources $p(V)$ (left panel), its $\lr{V}$ vs $\Na$ (2nd panel), ratio $\Na/(\bar{n} \lr{V})$ vs $\Na$ (3rd panel) and  ratio $\Na/(\bar{n} \lr{V})$ vs $\lr{V}$ (right panel) for different collision systems. The bottom panels show the same distributions but normalized by the knee, defined as $2A$ for left and right panels, and $2A\bar{n}$ for the middle two panels. The $\Na$ is generated with Par1 parameter set in Table~\ref{tab:1}.}
\end{figure}
\begin{figure}[h!]
\begin{center}
\includegraphics[width=0.8\linewidth]{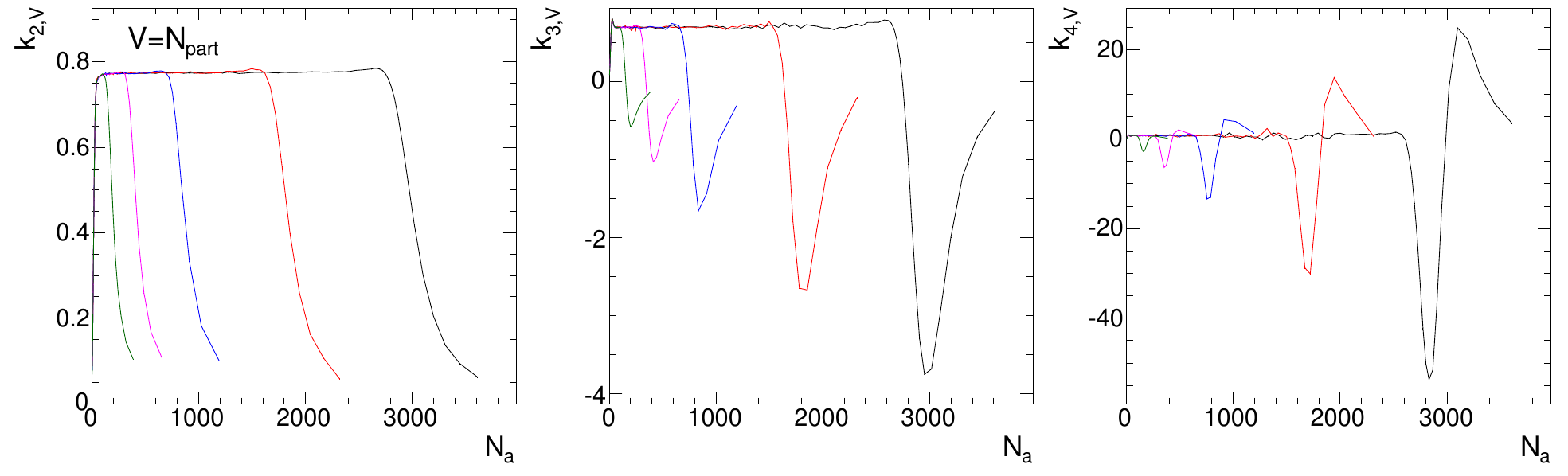}\\
\includegraphics[width=0.8\linewidth]{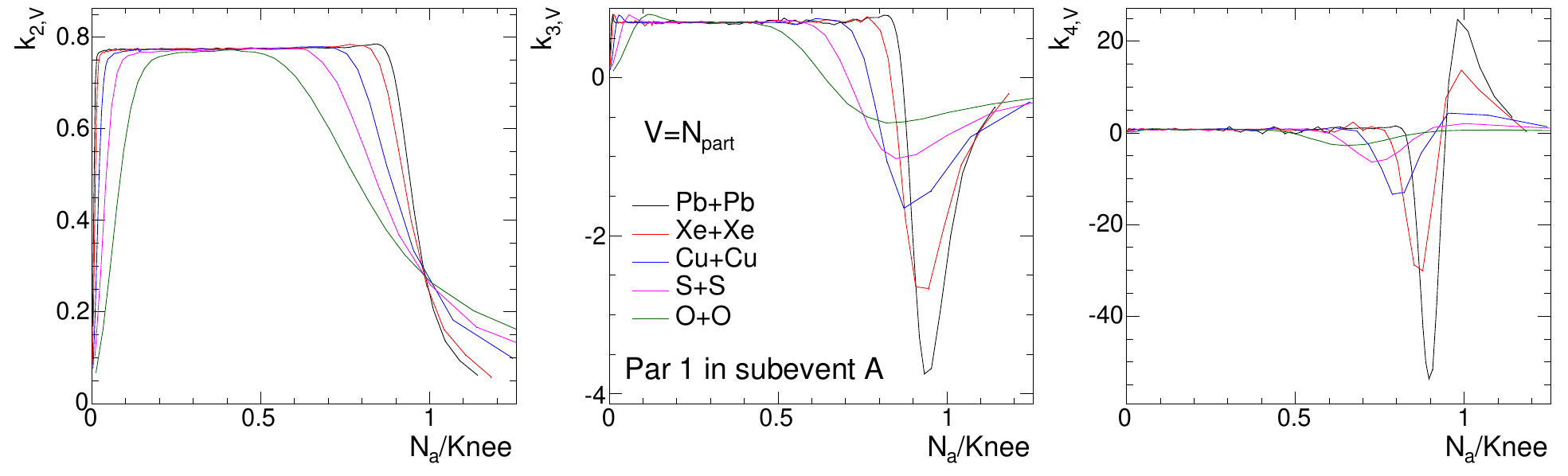}
\end{center}
\vspace*{-0.4cm}\caption{\label{fig:6} The centrality cumulants, $\kn{2}{\Ns}$ (left panel), $\kn{3}{\Ns}$ (middle panel),  and $\kn{4}{\Ns}$ (right panel) as a function $\Na$ (top row) or as a function of $\Na/N_{knee}$ (bottom row) for various collision systems. The $\Na$ is generated with Par1 parameter set in Table~\ref{tab:1}.}
\end{figure}
The centrality fluctuations are sensitive to the shape of $p(V\equiv\npart)$, which changes with the size of the collision system. The top-left panel of Fig.~\ref{fig:5} shows the distribution $p(V)$ from five collision systems in central collision region. The same distribution is replotted in the bottom-left panel, but the $x$-axes have been rescaled by $V^{\mathrm{max}}=2A$. The smaller system shows a broader tail in the $V$ distributions.  For system-size dependence studies, the centrality cumulants $k_{n,\Ns}$ are calculated with Par1 from Table~\ref{tab:1}. The results are presented both as a function of $\Na$ and as a function of $\Na/N_{\textrm{knee}}$, where the knee is defined as the average multiplicity for maximum $V=2A$ nucleons, $N_{\textrm{knee}}=2A\bar{n}$. The 2nd column of Fig.~\ref{fig:5} show the $k_{1,\Ns}=\lr{V}$ as a function of $\Na$ and $\Na/N_{\textrm{knee}}$. The $\lr{V}$ increases almost linearly with $\Na$, and flattens out close to $N_{\textrm{knee}}$. To quantify the non-linear behavior in the UCC region, we calculate the ratios $\Na/(\bara{n}\lr{\Ns})$ and present them in the third column. In the very small $\Na$ region,  the ratios are below unity and agree with each other, reflecting the fact that the $p(V)$ distributions have very similar shape in the small $V$ region. Towards the large $\Na$ region, the ratios separate from each other and increase to above unity with the smaller system showing a larger deviation from unity. The third bottom panel shows the ratios as a function of $\Na/N_{\textrm{knee}}$, where the increase in UCC is linear for all systems. For a smaller system, the rate of increase is smaller but over a broader range in $\Na/N_{\textrm{knee}}$, consistent with the smoother fall-off for smaller system in the central region shown of the bottom-left panel.

Previous studies on the particle production often focused on the scaling behavior of the particle production per source, i.e. $N/\lr{\Ns}$, with $\Ns$ defined as $\npart$ or the number of quark participants. They found that the sources based on the quark participant number has a better scaling behavior than that based on the nucleon participant number from $pp$, $\pA$ to A+A systems~\cite{Adler:2013aqf,Lacey:2016hqy,Loizides:2016djv,Bozek:2016kpf,Acharya:2018hhy}. However, a detailed study by ALICE Collaboration showed that both types sources show a sharp increase of $N/\lr{\Ns}$ vs $\lr{\Ns}$ in the UCC region~\cite{Acharya:2018hhy}. To perform a similar check, the right column of Fig.~\ref{fig:5} shows $\Na/(\bara{n}\lr{\Ns})$ as a function of $\lr{\Ns}$ and $\lr{\Ns}/2A$, where $\lr{\Ns}$ is obtained by mapping from $\Na$ using the data in the top panel in the second column. The linear increase in the third column now appears as a sharp increase in the UCC region, similar to the data. The increase is stronger and affects most of the centrality range in a small system, consistent with the experimental observation.

On the other hand, the experimental observable is defined as $\Nb/\lr{\Ns}$ as appose to $\Na/\lr{\Ns}$ in our study. In the independent source model framework, $\Nb/\lr{\Ns}$ would always equal to $\bar{n}$ by construction. Therefore, one natural explanation would be that there are strong correlations between the particle production in subevents A and B, i.e. $\na$ and $\nb$ are strongly correlated. In an extreme case of $\na=\nb$, the $\Nb=\Na$ would be valid for each event, and $\Nb/\lr{\Ns}$ would also show an increase in the UCC region, we leave this point to a future investigation.

Figure~\ref{fig:6} shows higher-order $k_{n,\Ns}$ as a function of $\Na$ (top row) or $\Na/N_{\textrm{knee}}$ (bottom row). The $k_{n,\Ns}$ have the same constant values in mid-central collisions, but deviates from each other towards more peripheral or more central region where the lower and upper boundary effects from the $p(V)$ distribution are important. Such deviation appears over a larger fraction of the $\Na$ range for smaller collision systems. In the central region, the magnitudes of the $k_{3,\Ns}$ and $k_{4,\Ns}$ are smaller for smaller collision systems. At the same time, the widths of the maxima region are larger in terms of $\Na/N_{\textrm{knee}}$, implying that the centrality fluctuations influences a larger fraction of the centrality region. This can be understood from the bottom-left panel in Fig.~\ref{fig:5} , which shows that the decrease in the central region is smoother in smaller collision systems.

\section{Results on forward-backward centrality decorrelations}\label{sec:6}

To study the effects of centrality decorrelations, we consider $V_{a}=\npartf$ and $V_{b}=\npartb$ as the sources for particles in subevent A and B, respectively. According to Eq.~\eqref{eq:8}, the cumulants of $p(V_b)$ are expressed in terms of cumulants $k_{n,ab}$ and $k_{n,ab}'$ describing $p(\Nsb)_{\Nsa}$, as well as $k_{n,\Nsa}$ describing the fluctuation of $V_{a}$. In the following, we first discuss the behaviors of $k_{n,\Nsa}$, $k_{n,ab}$ and $k_{n,ab}'$ and then show quantitatively how well $k_{n,\Nsb}$ are described by Eq.~\eqref{eq:8}.

\begin{figure}[h!]
\begin{center}
\includegraphics[width=1\linewidth]{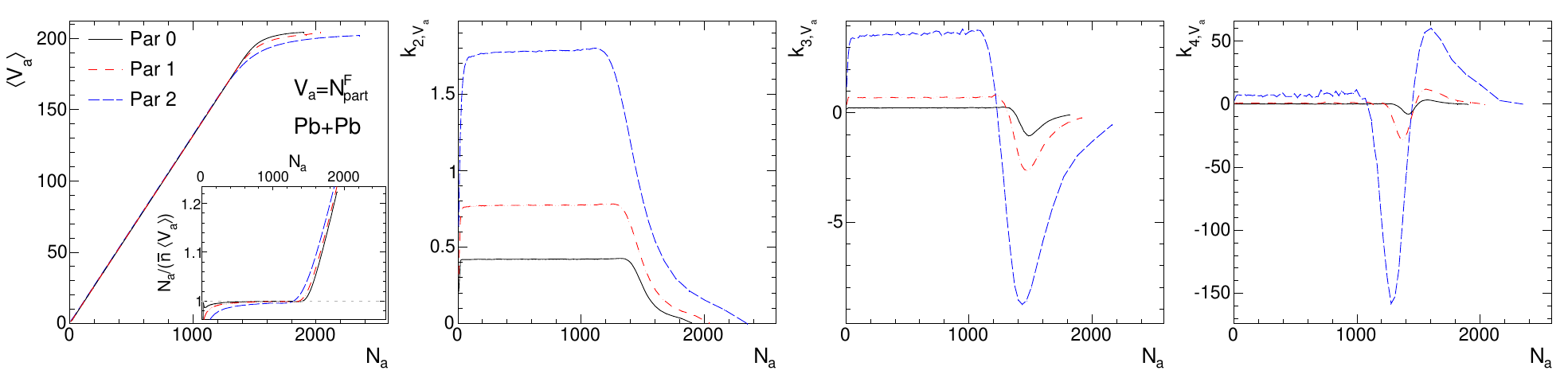}\\
\includegraphics[width=1\linewidth]{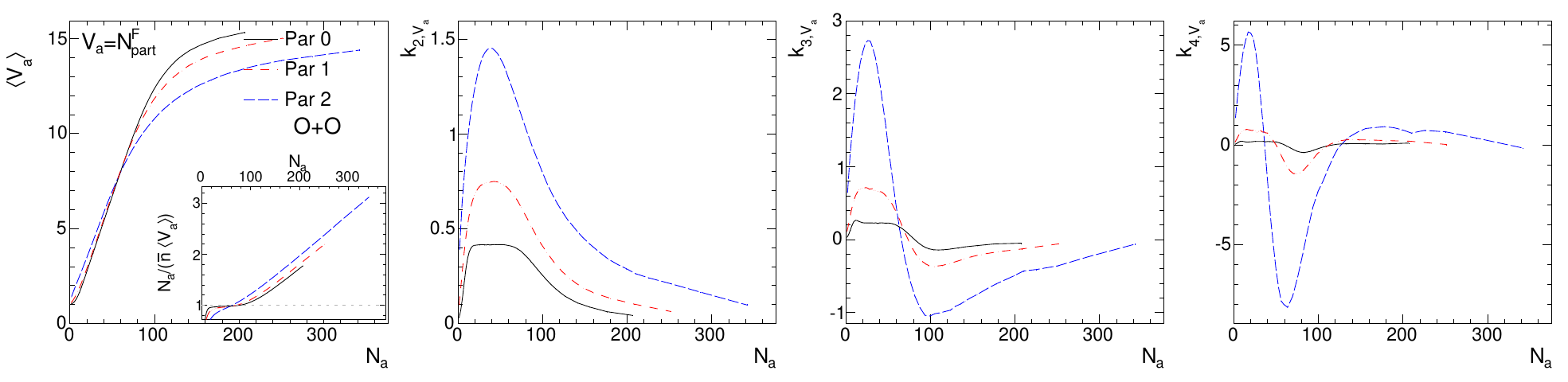}
\end{center}
\vspace*{-0.5cm}\caption{\label{fig:7} The centrality cumulants for $p(\Nsa)$ with $\Nsa\equiv\npartf$, $k_{1,\Nsa}=\lr{\Nsa}$ (left column),  $k_{2,\Nsa}$ (2nd column),  $k_{3,\Nsa}$ (3rd column), $k_{4,\Nsa}$ (right column) as a function of $\Na$ from Pb+Pb system (top row) and O+O system (bottom row). The insert panels show the ratios $\Na/(\bar{n} \lr{\Nsa})$ to quantify the deviation of $\Na$ dependence of $\lr{\Nsa}$ from the linear relations.}
\end{figure}

Figure~\ref{fig:7} shows the centrality cumulants $k_{n,\Nsa}$ for Pb+Pb and O+O collisions. The magnitudes of $k_{n,\Nsa}$ depend on the centrality resolution of subevent A, and are calculated with Par0--Par2 as a function of $\Na$. The results for Pb+Pb are similar to those presented in the bottom row of Fig.~\ref{fig:3}. The only difference is that the $\npartf$ instead of $\npart$ is used as sources in this figure, therefore the maximum range of $\Na$ reaches only about half of that in Fig.~\ref{fig:3}. The results for O+O are much more affected by the boundary effects, such that no plateau is observed for the second- and higher-order cumulants. 

\begin{figure}[h!]
\begin{center}
\includegraphics[width=0.9\linewidth]{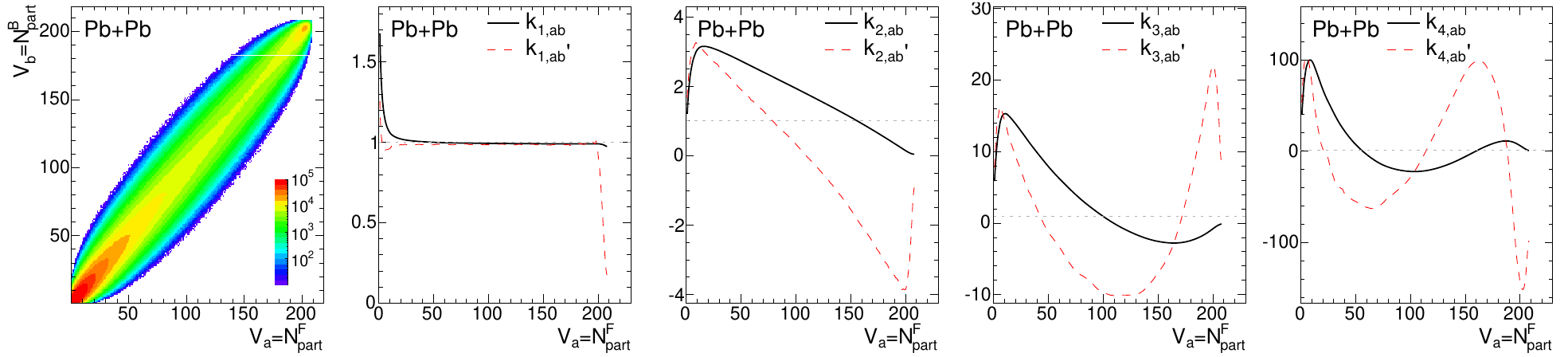}\\
\includegraphics[width=0.9\linewidth]{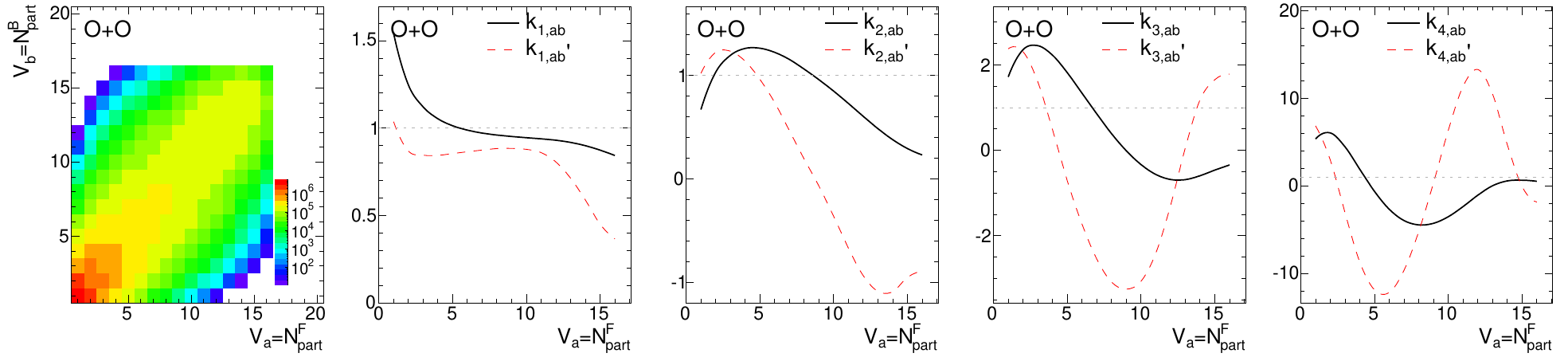}
\end{center}
\vspace*{-0.5cm}\caption{\label{fig:8} The 2D correlation between $\Nsa=\npartf$ and $\Nsb=\npartb$ (left column) and corresponding $k_{n,ab}$ and $k_{n,ab}'$ for $n=1$ (2nd column), $n=2$ (3rd column),$n=3$ (4th column) and $n=4$ (right column) in Pb+Pb system (top row) and O+O system (bottom row).}
\end{figure}

Next we discuss the behavior of $k_{n,ab}$ and $k_{n,ab}'$ describing $p(\Nsb)_{\Nsa}$. The latter is rather spread out as shown in the left panels of Fig.~\ref{fig:8} for Pb+Pb and O+O collisions. In principle, $p(\Nsb)_{\Nsa}$ is not necessarily described by independent source picture, i.e. the value of $\Nsb$ can not be treated as a sum of independent contributions from $\Nsa$ number of sources. In practice, the deviation from the independent source picture is small and the cumulants of $p(\Nsb)_{\Nsa}$ are approximately proportional to $\Nsa$.  Nevertheless, we need to use Eq.~\eqref{eq:9} to calculate $k_{n,ab}'$, which includes the first-order correction to account for the residual dependence of $k_{n,ab}$ on $\Nsa$. Such corrections can be quite important if the centrality resolution of subevent A is poor (i.e. $\kn{n}{\Nsa}$ are large). 

The right panels of Fig.~\ref{fig:8} show $k_{n,ab}$ and $k_{n,ab}'$ as a function of $\Nsa$. The first-order cumulants $k_{1,ab}$ and $k_{1,ab}'$ are close to unity in mid-central collisions, where $\lr{\Nsb}_{\Nsa}\approx \Nsa$. The rather sharp decrease of $k_{n,ab}'$ in central region is due to $\Nsa \frac{\partial k_{1,ab}}{\partial \Nsa }$ in Eq.~\eqref{eq:9}, which are quite large in the central region. Overall the influence of $k_{1,ab}$ and $k_{1,ab}'$ to the higher-order $\kn{n}{\Nsb}$ should be small except in the very central and peripheral collisions.

The third column of Fig.~\ref{fig:8} shows that the scaled variance $k_{2,ab}$ increases sharply and then decreases gradually for the remaining $\Nsa$ range. This behavior can be understood from the leaf-like shape of the 2D correlation between $\Nsb$ and $\Nsa$: although the variance of $\Nsb$ is small towards small and large $\Nsa$ region, the scaled variance of $\Nsb$ (i.e. normalized by $\lr{\Nsb}$) is large in small $\Nsa$ region and decreases to nearly zero at largest $\Nsa$. The higher-order cumulants $k_{3,ab}$ and  $k_{4,ab}$ have more complex behaviors. In the small $\Nsa$ region,  $k_{n,ab}$ are  positive since $\Nsb>0$. At large $\Nsa$, the fluctuations of $\Nsb$ is bounded by $\Nsb\leq A$, leading to a negative $k_{n,ab}$. Note that the sign-change of cumulants occurs in mid-central region, around $\Nsa\approx 110$ for $k_{3,ab}$ and $\Nsa\approx 60$ for $k_{4,ab}$ for Pb+Pb. This behavior is very different from the influence of pure multiplicity fluctuations in Section~\ref{sec:5}, where the sign-change happens only in the UCC region. For higher-order cumulants, we note that the value of $k_{n,ab}'$ are very different from $k_{n,ab}$ over the full range of $\Nsa$ due to $\Nsa \frac{\partial (k_{1,ab}k_{n,ab})}{k_{1,ab} \partial \Nsa }$ in Eq.~\eqref{eq:9}. This implies that the deviation from the independent source assumption has a stronger impact for the third and higher-order cumulants.

In the smaller O+O system, even the $k_{1,ab}$ and $k_{1,ab}'$ already show a quite sizable deviation from unity. In general, the higher-order cumulants $k_{n,ab}$ and $k_{n,ab}'$ have smoother variations but over a much broader $\Nsa$ range.

From $k_{n,\Nsa}$ in Fig.~\ref{fig:7} and $k_{n,ab}$ and $k_{n,ab}'$ in Fig.~\ref{fig:8}, we calculate the $k_{n,\Nsb}$ via Eq.~\eqref{eq:8} and compare with the results from full calculation. This comparison is shown in Fig.~\ref{fig:9} for Pb+Pb and O+O collisions, together with the breakdown of the contributions from its individual components. The behavior of cumulants and the comparison can be summarized as follows:
\begin{figure}[h!]
\begin{center}
\includegraphics[width=0.8\linewidth]{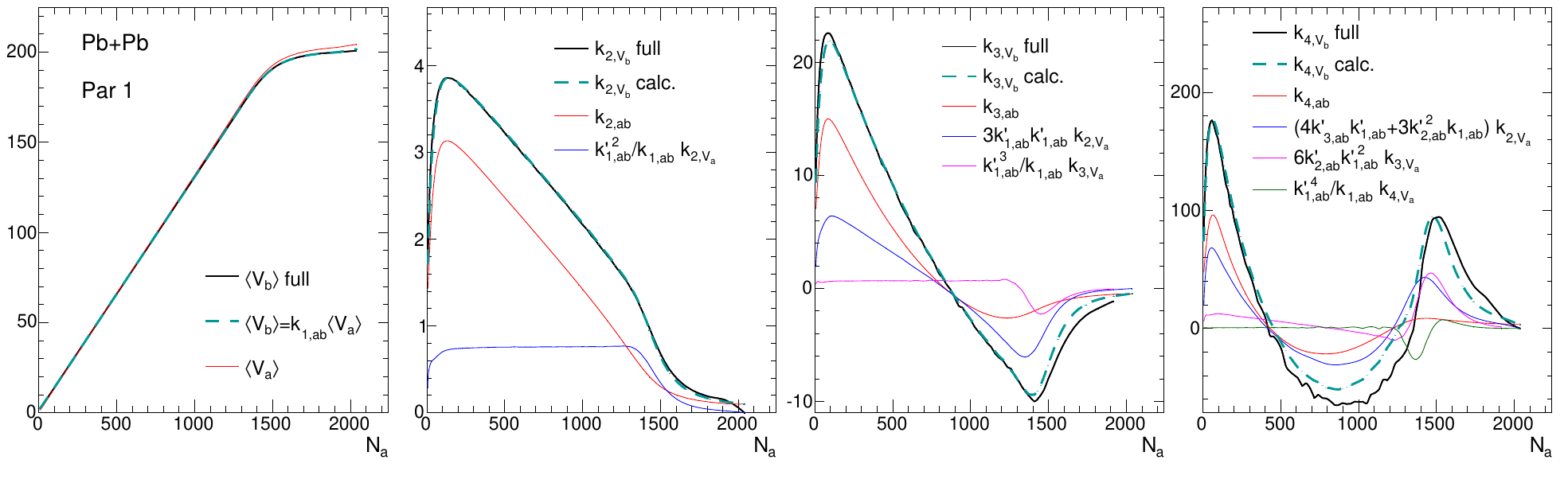}\\
\includegraphics[width=0.8\linewidth]{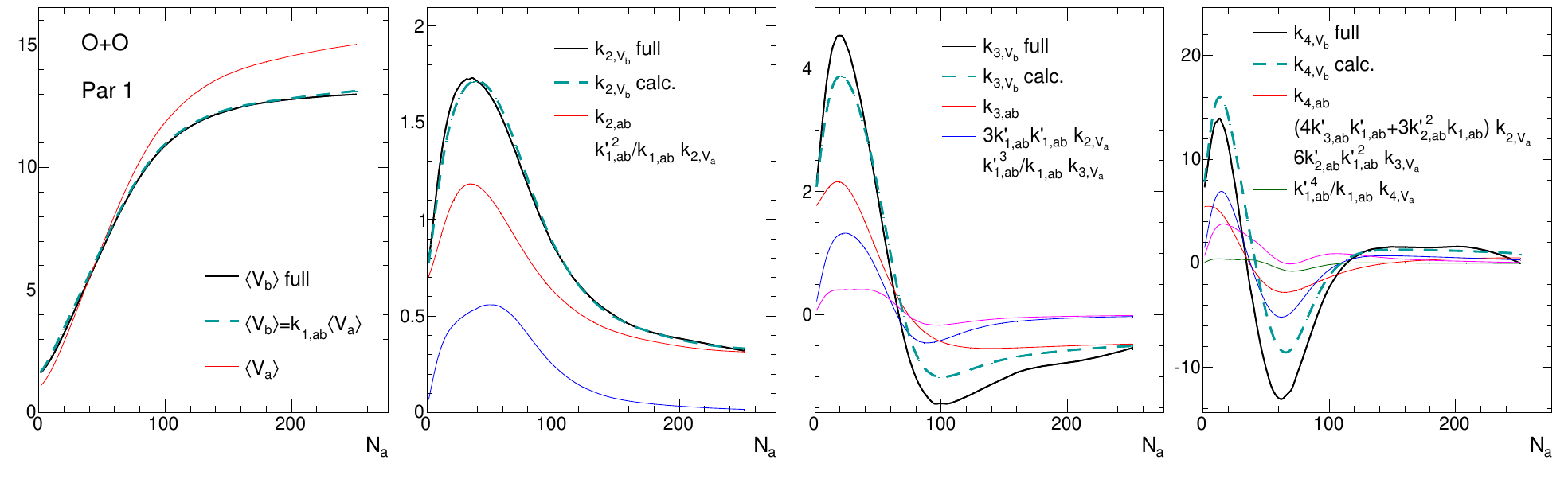}
\end{center}
\vspace*{-0.5cm}
\caption{\label{fig:9} The centrality cumulants $k_{n,\Nsb}$ in subevent B as a function of $\Na$, obtained from full calculation (thick solid line) or calculation via Eq.\eqref{eq:11} (thick dashed lines) and its various components (thin solid lines) for $n=1$ (the first column), $n=2$ (2nd column), $n=3$ (3rd column) and $n=4$ (right column) in the Pb+Pb system (top row) and the O+O system (bottom row).}
\end{figure}

\begin{enumerate}
\item The $\lr{\Nsb}_{\Nsa}$ in the left panels agrees very well with Eq.~\eqref{eq:8}. In Pb+Pb collisions, the $\lr{\Nsb}_{\Nsa}$ is nearly identical to $\lr{\Nsa}$, but significant deviation is observed in O+O collisions, which can be fully explained by the behavior of $k_{1,ab}$ shown in Fig.~\ref{fig:8}. The flattening behavior of $\lr{\Nsb}_{\Nsa}$ in central collisions is expected to be dominated by pure multiplicity smearing effects in subevent A, and the centrality decorrelations is important only for small collision systems. 
 
\item The results of $k_{2,\Nsb}$ are shown in the second column. The $\Na$ dependence of $k_{2,\Nsb}$ in Pb+Pb collisions is described nearly perfectly by Eq.~\eqref{eq:8}, while a small deviation is observed in O+O collisions around the maximum of $k_{2,\Nsb}$. In Pb+Pb collisions, the decrease of $k_{2,\Nsb}$ as a function of $\Na$ is mainly driven by $\kn{2}{ab}$, while $k_{2,\Nsa}$ mainly contributes an offset in mid-central region and a sharp decrease in the UCC region. In O+O collisions, these two components have more similar shapes, although the $\kn{2}{ab}$ term dominates in the central region and the most peripheral region. 

\item The results of $k_{3,\Nsb}$ are shown in the third column. In Pb+Pb collisions, the description by Eq.~\eqref{eq:8} is also very good, except that it slightly underestimates the magnitude in the most central collisions. In O+O collisions, Eq.~\eqref{eq:8} underestimates the magnitude of $k_{3,\Nsb}$ in the central region, and around its maximum in the peripheral region. Looking at the three individual components of Eq.~\eqref{eq:8}, the genuine FB decorrelation term $\kn{3}{ab}$ dominates only in the peripheral region, but its contribution is not as important as the second term associated with $k_{2,\Nsa|\Na}$. The last term associated with $k_{3,\Nsa|\Na}$ is nearly a constant, and it has little influence on the shape of $k_{3,\Nsb}$, except in central collisions. These behaviors also imply that $\kn{3}{ab}$ can only be reliably extracted if $k_{3,\Nsa|\Na}$ are small; otherwise, one has to first measure $\kn{2}{ab}'$ and $\kn{1}{ab}'$ using the information in the first two columns.

\item The behavior of $k_{4,\Nsb}$ shown in the right column are much more complex. In Pb+Pb collisions, the description of $k_{4,\Nsb}$ by Eq.~\eqref{eq:8} is only good in peripheral and the most central collisions. In O+O collisions, Eq.~\eqref{eq:8} has poor description over most of the region for the Par1 parameter set. When the Par0 parameter set is used, the agreement is much better in Pb+Pb collisions, but is still relatively poor in O+O collisions. Looking at the contributions of individual terms, the intrinsic kurtosis term $k_{4,ab}$ from FB fluctuations is important only in the very peripheral and central regions. The mixing terms between lower-order $k_{n,ab}$ and $k_{n,\Nsa}$ dominate in the mid-central collisions and are also important in the other centrality range. 
\end{enumerate}

\begin{figure}[h!]
\begin{center}
\includegraphics[width=0.8\linewidth]{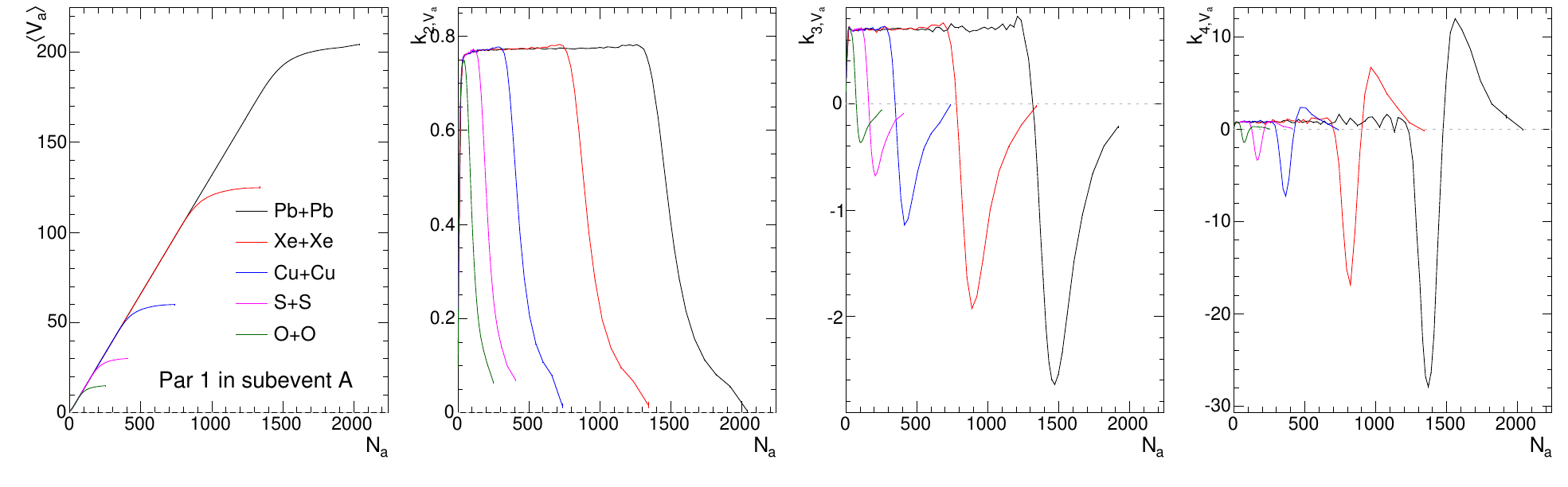}\\
\includegraphics[width=0.8\linewidth]{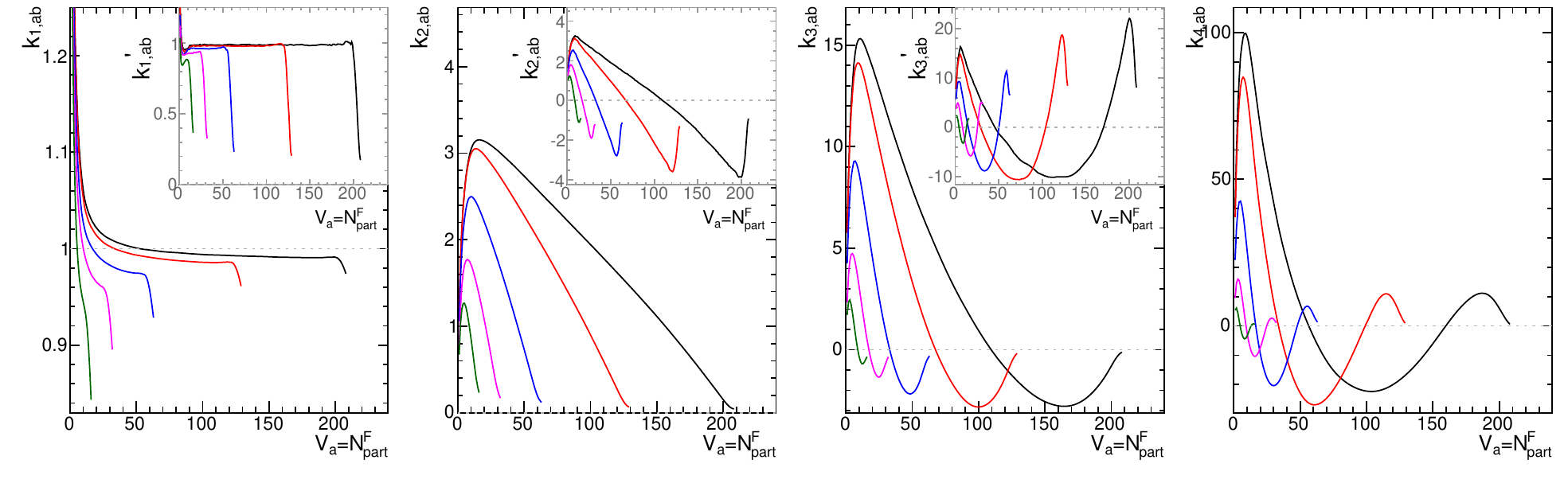}\\
\includegraphics[width=0.8\linewidth]{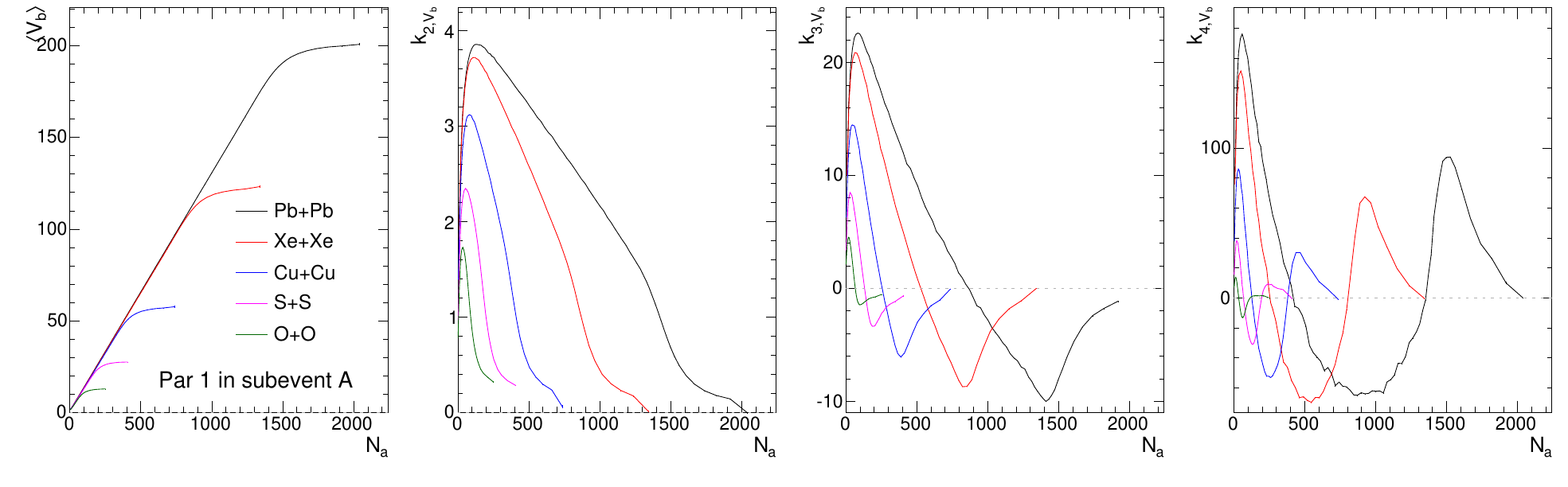}
\end{center}
\vspace*{-0.5cm}
\caption{\label{fig:10}  The centrality cumulants in subevent A $k_{n,\Nsa}$ as a function of $\Na$ (top row), the cumulants for forward-backward source correlation, $k_{n,ab}$ and  $k_{n,ab}'$, as a function of $\Nsa$ (middle row), and the centrality cumulants in subevent B $k_{n,\Nsb}$ as a function of $\Na$ (bottom row) for various collision systems, for $n=1$ (first column), $n=2$ (2nd column), $n=3$ (3rd column) and $n=4$ (right column).}
\end{figure}

Figure~\ref{fig:10} summarizes the results in the five collision systems for Par1 parameter set. The results of $k_{n,\Nsb}$ in the bottom row are related to $k_{n,\Nsa}$ in the top row and $k_{n,ab}$ and $k_{n,ab}'$ in the middle row via Eq.~\eqref{eq:8}. For the first-order cumulants shown in the left column, the behavior of $\lr{\Nsb}_{\Nsa}$ largely resembles those of $\lr{\Nsa}$, except for very small systems where the $k_{1,ab}$ show a significant deviation from one. For all higher-order cumulants, the values of $k_{n,\Nsa}$ are much smaller than $k_{n,ab}$.  In fact, in the absence of the centrality resolution effect in subevent A, $k_{n,\Nsa}$ vanish and $k_{n,\Nsb}$ approach $k_{n,ab}$. We find that if Par2 is chosen for multiplicity smearing in subevent A, the $k_{n,\Nsa}$ are much larger and dominate the behavior of $k_{n,\Nsb}$. The shapes of $k_{n,ab}$ and $k_{n,ab}'$ are similar between different collision systems, but the magnitudes are smaller for smaller collision systems. This leads to a similar system-size ordering for the values of $k_{n,\Nsb}$. 

Given the fact that $k_{1,ab}$ and $k_{1,ab}'$ are close to one, the features of $k_{2,\Nsb}$ can be used to separate $k_{2,ab}$ and $k_{2,\Nsa}$ via Eq.~\eqref{eq:8}: the increase and decrease from peripheral to mid-central collisions mainly reflects the contribution of $k_{2,ab}$, while the sharp decrease and flattening out behavior in central region are dominated by the $k_{2,\Nsa}$. In contrast, we find that $k_{3,\Nsb}$ and $k_{4,\Nsb}$ are much more sensitive to mixing terms between $k_{n,ab}$, $k_{n,ab}'$ and $k_{n,\Nsa}$, especially in mid-central and central regions, similar to Fig.~\ref{fig:9}. Separate these different components would be a rather challenging task.

In the final step, the centrality cumulants $k_{n,\Nsb}$ shown in Figs.~\ref{fig:9} and ~\ref{fig:10} are used to obtain the full results of multiplicity cumulants including the smearing effects in subevent B via Eq.~\eqref{eq:5} or Eq.~\eqref{eq:11} for Par1.  Examples can be found in Sec.~\ref{sec:app2}.

\section{Summary and discussion}\label{sec:7}
In heavy ion collisions, the centrality or the volume of the fireball, characterized by the number of sources $\Ns$ in the initial state geometry, is not fixed but fluctuates for events with the same final-state particle multiplicity $N$. This so-called centrality or volume fluctuations (CF) can arise from either 1) fluctuations in the particle production process which smears the mapping between $N$ and $\Ns$ or 2) longitudinal fluctuations of $\Ns$ within the same event which decorrelates the $N$ between different $\eta$ ranges. In this paper, we propose to study CF using the correlation of multiplicities $\Nb$ and $\Na$ in two subevents separated in pseudorapidity. This two-dimensional correlation are analyzed via cumulants of the conditional probability distribution $p(\Nb)_{\Na}$ as a function of $\Na$. The contributions from the two types of CFs can be identified from features in the $\Na$ dependence of these cumulants.

For a quantitative study of the CF, a standard Glauber model based on an independent source picture is used: the Glauber model is used to produce the $\Ns$ for each event, and the final-state particle multiplicity in each event is then calculated as a sum of particles from each source $N=\sum_{i=1}^{\Ns} n_i$ with $n_i$ sampled independently from a common $p(n)$. For the study on the effects of multiplicity smearings, the sources for subevents A and B are chosen as $\Nsa=\Nsb=\npart$. For the study on the effects of centrality decorrelations, they are chosen as $\Nsa=\npartf$ and $\Nsb=\npartb$. The centrality selection is based on $\Na$ for both cases, i.e. cumulants are calculated upto fourth-order:  mean $\lr{\Ns}$, scaled variance $k_{2,\Ns}=\lr{(\delta\Ns)^2}/\lr{\Ns}$, skewness $k_{3,\Ns}=\lr{(\delta\Ns)^3}/\lr{\Ns}$ and kurtosis $k_{4,\Ns}=(\lr{(\delta\Ns^4)}-3\lr{(\delta\Ns^2)}^{_2})/\lr{\Ns}$.

The $\lr{\Ns}$ is observed to be linearly proportional to $\Na$, except in ultra-central collisions (UCC) where the increase slows down as $\Na$ approaches the upper bound. This means that $\Na/\lr{\Ns}$ exhibits a sharp increase in the UCC region similar to what is observed in the experimental data~\cite{Acharya:2018hhy,Aaboud:2019sma}. We also verified that (not shown) this linearity is not affected much by the centrality decorrelations considered in this paper, except in smaller systems where the correlation between $\Nsa$ and $\Nsb$ has significant deviations from diagonal. Such behavior in the UCC is very sensitive to the properties of $p(\Ns)$ and $p(n)$. 

When including only effects of multiplicity smearings, the scaled variance $k_{2,\Ns}$ is nearly constant in mid-central collisions, but decreases to zero in UCC events as $\Na$ approaches the upper bound. In the presence of centrality decorrelations, the $k_{2,\Ns}$ also exhibits a linear decrease in the mid-central collisions. The rate of decrease should be a robust measure of the extent of longitudinal fluctuations in the particle production sources. 

The $\Na$ dependence of higher-order centrality cumulants are more complex. In the presence of centrality decorrelations, they receive significant contributions from several mixing terms related to the lower-order cumulants, which are also much larger in smaller systems. Furthermore, additional correction terms need to be included to account for the fact that the cumulants for $p(\Nsb)_{\Nsa}$ is not proportional to $\Nsa$ except for the first order. Due to these reasons, the $\lr{\Ns}$ and $k_{2,\Ns}$ are considered most valuable observables to probe the nature of $p(\Ns)$ and $p(n)$. 

From the centrality cumulants $k_{n,\Ns}$, we calculated the cumulants for final-state multiplicity $\Nb$ in subevent B, $K_{n,\mathrm{b}}$ via Eq.~\eqref{eq:5}. We find that the flattening behavior of $\lr{\Nb}$ as a function of $\Na$ in UCC remains. Experimentally, one can calculate both $\lr{\Nb} (\Na)$ and $\lr{\Na} (\Nb)$ from $p(\Nb,\Na)$ to quantify the relative centrality resolution of the two subevents. If $\lr{\Nb} (\Na)$ is more diagonal than $\lr{\Na} (\Nb)$, it would imply that $\Na$ has a better centrality resolution on $\Ns$ than $\Nb$ (a similar conclusion was reached from ATLAS data~\cite{Aaboud:2019sma}). For the scaled variance $K_{2,\mathrm{b}}$, we find that most of the features of $k_{2,\Ns}$ has been preserved, therefore experimentally measured scaled variance in subevent B for events selected in subevent A can also be used to extract information about the centrality decorrelations. 
 
The centrality decorrelations considered in this paper is rather simplistic. This can be improved by considering the partonic degree of freedom in the nucleons and assuming the quark participant number $n_{qp}$ to be a number that vary with $\eta$. In addition, one should consider situation used by most collider experiments, where the subevent A for centrality is defined in both forward and backward rapidity, and subevent B is defined at mid-rapidity. In this case, although $\Nsa$ and $\Nsb$ are largely correlated with $\npart$, they should still be decorrelated due to difference in $n_{qp}$. We leave this to future investigations.

In summary, we have studied the centrality fluctuations in terms of the number of initial sources $\Ns$ using the correlation of the final-state multiplicity between two subevents $\Na$ and $\Nb$. The sources and final multiplicity are generated with a Glauber model within an independent source picture. This correlation is analyzed in terms of cumulants for multiplicity distribution in one subevent ($\Nb$) for events selected in another subevent with a fixed multiplicity ($\Na$). The $\Na$ dependence of the cumulants are affected by both pure multiplicity smearing in the particle production as well as the decorrelation between the sources in the two subevents. We find that the behavior of cumulants in ultra-central collisions are very sensitive to the particle production for each source $p(n)$ and $p(\Ns)$ due to the steeply falling $p(\Ns)$ distribution. The FB centrality decorrelations have very little impact on the mean multiplicity but leads to an decrease of scaled variance $\lr{(\delta\Nsb)^2}/\lr{\Nsb}$ and $\lr{(\delta\Nb)^2}/\lr{\Nb}$ as a function of $\Na$. These features can be used to constrain the particle production mechanism as well as the longitudinal fluctuations of the initial-state sources.

This research is supported by National Science Foundation under grant number PHY-1613294 and PHY-1913138, and the National Natural Science Foundation of China under grant number 11922514.
\appendix

\section{Cumulants for correlated variables}\label{sec:app1}
The cumulants for a given probability density function (pdf) $p(x)$, $c_{1,x} = \bar{x}, c_{2,x}=\lr{(\delta x)^2}, c_{3,x}=\lr{(\delta x)^3},...$, are defined from the cumulant generating function $\chi_x(t)$:
\begin{align}\label{app:1}
\chi_x(t) = \ln \int dx p(x)e^{xt}\equiv \ln \lr{e^{xt}}=\sum_n c_{n,x} \frac{t^n}{n!}.
\end{align}
Let's consider a quantity $B$, whose value is related to the size of the system described by the number of sources $V$,
\begin{align}\label{app:2}
p(B)= \int dV p(B)_{V}p(V).
\end{align}
The $p(B)_{V}$ is the distribution of $B$ for fixed value of $V$. There are three pdfs $p(B), p(B)_{V}$ and $p(V)$ with their own cumulant generating functions:
\begin{align}\label{app:3}
&\chi_B (t) = \ln \lr{e^{Bt}},\\
&\chi_{B|V} (t) = \ln \lr{e^{Bt}}_{V},\\
&\chi_V (t) = \ln \lr{e^{Vt}}.
\end{align}

In the independent source picture, the value of $B$ in each event is given by $B=\sum_{i=1}^{V} b_i$, with $b_i$ generated independently from a common pdf $p(b)$. The additivity of cumulants implies that, $\chi_{B|V}(t) = V \times \chi_b (t)$, where 
\begin{align}\label{app:4}
&\chi_b (t) = \ln \lr{e^{bt}}
\end{align}
is the cumulant generating function for $p(b)$. Therefore,
\begin{align}\label{app:5}
&\chi_B (t) = \ln \lr{e^{Bt}}= \ln \lr{\lr{e^{Bt}}_V} = \ln \lr{e^{V\chi_b(t)}}= \sum_n c_{n,V} \frac{\left(\chi_b(t)\right)^n}{n!}= \sum_{n,m} c_{n,V} \frac{\left(c_{m,b} t^m/m!\right)^n}{n!}.
\end{align}
Collect terms in power of $t$ on both sides, one obtains the general formula of Ref.~\cite{Skokov:2012ds}, which expresses the cumulants of $p(B)$ in terms of those for $p(b)$ and $p(V)$, i.e.
\begin{align}\label{app:6}
 c_{n,B}=\sum_{i=1}^{n}  c_{i,V} B_{n,i}(c_{1,b},c_{2,b},..., c_{n-i+1,b}),
\end{align}
where the $B_{n,i}$ are Bell polynomials.

This paper considers deviations from the independent source picture, i.e. when $\chi_{B|V}$ is not exactly proportional to $V$,
\begin{align}\nonumber
\chi_{B|V}(t,V) &= V  \chi_b (t,V) =V\left(\chi_b(t,\bar{V})+\sum_{n=1}^{\infty}\chi_b^{(n)}(t,\bar{V}) \frac{(\delta V/\bar{V})^n}{n!}\right)\\\label{app:7}
&=\bar{V}\chi_b+ \delta V\left(\chi_b+\chi_b^{(1)}\right)+ \frac{1}{\bar{V}}(\delta V)^2\left[\chi_b^{(1)}+\chi_b^{(2)}/2\right]+...+\frac{1}{\bar{V}^n} (\delta V)^{n+1}\left[\chi_b^{(n)}/n!+\chi_b^{(n+1)}/(n+1)!\right]+...,
\end{align}
where we has Taylor-expanded $\chi_b$ in $\ln V$ around $\bar{V}$, i.e.,
\begin{align}\label{app:8}
\chi_b^{(n)}(t) = \frac{\partial^n\chi_b(t,V)}{(\partial \ln V)^n}|_{V=\bar{V}}\;.
\end{align}
Dropping terms that are suppressed by the system size $1/\bar{V}^n$, and plugging the first two terms back into Eq.~\eqref{app:5} gives
\begin{align}\label{app:9}
&\chi_B (t)= \ln \lr{e^{V\chi_b(t,V)}} \approx \ln \lr{e^{ V \left(\chi_b(t)+\chi_b^{(1)}(t)\right)}} - \bar{V}\chi_b^{(1)}(t)
\end{align}
Following the remaining steps of Eq.~\eqref{app:5}, we obtain:
\begin{align}\label{app:10}
 c_{n,B}= c_{1,V}c_{n,b}+ \sum_{i=2}^{n}  c_{i,V} B_{n,i}(\frac{\partial (c_{1,b}V)}{\partial V},\frac{\partial (c_{2,b}V)}{\partial V},..., \frac{\partial (c_{n-i+1,b}V)}{\partial V})
\end{align}
This result can also be expressed with the normalized cumulant notation, i.e.,
\begin{align}\nonumber
k_{1,B}&= k_{1,V}k_{1,b}\;\; \mathrm{or}\;\; \bar{B} = \bar{V}\bar{b},\\\label{app:11}
k_{n,B}&= k_{n,b}+ \sum_{i=2}^{n}  k_{i,V}  B_{n,i}(\frac{\partial (k_{1,b}V)}{\partial V},\frac{\partial (k_{1,b}k_{2,b}V)}{k_{1,b}\partial V},..., \frac{\partial (k_{1,b}k_{n-i+1,b}V)}{k_{1,b}\partial V}).
\end{align}
For large collision systems, this leading-order approximation works very well. For small collision systems, where the variance of CF may not be small in comparison to $\bar{V}$,  higher-order correction terms in Eq.~\eqref{app:7} should also be considered. The expressions of the corrections for $c_{n,B}$ are quite lengthy, so we only show the result for first and second cumulants up to all order, and the third-order cumulants to $1/\bar{V}$,
\begin{align}\nonumber
\delta c_{1,B}&= \sum_{k=1}^{\infty}\frac{1}{\bar{V}^k}\lr{(\delta V)^{1+k}} D^{(k)}_{1}\approx\frac{1}{\bar{V}} c_{2,V} D^{(1)}_{1}\;,\\\nonumber
\delta c_{2,B}&= \sum_{k=1}^{\infty}\frac{1}{\bar{V}^k}\left(\lr{(\delta V)^{1+k}} D^{(k)}_{2}+2\lr{(\delta V)^{2+k}}D^{(0)}_{1} D^{(k)}_{1}\right)\approx\frac{1}{\bar{V}}\left(c_{2,V}  D^{(1)}_{2}+2c_{3,V} D^{(0)}_{1} D^{(1)}_{1}\right)\;,\\\label{app:11b}
\delta c_{3,B}&\approx \frac{1}{\bar{V}}\left(c_{2,V}  D^{(1)}_{3}+3c_{3,V} (D^{(0)}_{1} D^{(1)}_{2}+D^{(0)}_{2} D^{(1)}_{1})+3(c_{4,V}+4c_{2,V}^2)D^{(0)}_{1}D^{(0)}_{1}D^{(1)}_{1}\right)\;,
\end{align}
with the short-hand notation $D^{(k)}_{n}\equiv c_{n,b}^{(k)}/k!+c_{n,b}^{(k+1)}/(k+1)!$. 

As a concrete application of Eqs.~\eqref{app:10}-\ref{app:11}, it is useful to consider the CF due to the finite centrality bin-width effect~\cite{Luo:2013bmi}. Assuming centrality is defined in a finite multiplicity range in subevent A, whose distribution is described by $p(A)$, the corresponding CF is given by:
\begin{align}\label{app:12}
p(V)= \int dA p(V)_{A}p(A).
\end{align}
In general, $p(V)_{A}$ is not described by the independent source picture, i.e. the number of $V$ per particle is not independent, even if $p(A)_{V}$ is. But the correlation between $V$ and $A$ is expected to be close to linear, and thus the approximation described by Eq.~\eqref{app:10} or \ref{app:11} can be directly used. 

When one needs to consider both the centrality fluctuation in subevent B and the centrality bin width effect in subevent $A$, the relations can be expressed as,
\begin{align}\nonumber
&p(B) = \int dV p(B)_{V}p(V),\\\label{app:13}
&p(V) = \int dV p(V)_{A}p(A)
\end{align}
We consider the probability distribution $p(v)$ of $V$ per particle, whose cumulant generating function $\chi_v (t)$ is related to those for $p(V)_{A}$ as
\begin{align}\label{app:14}
\chi_{V|A}(t,A) &= A  \chi_v (t,A).
\end{align}
Combining Eqs.\ref{app:6} and \ref{app:10}, we obtain 
\begin{align}\nonumber
&\bar{B} = \bar{A}\bar{v}\bar{b},\\\label{app:15}
& c_{n,B}=\sum_{i=2}^{n} \left(c_{1,A}c_{j,v}+ \sum_{j=2}^{i}  c_{j,A} B_{i,j}\left(\frac{\partial (c_{1,v}A)}{\partial A},\frac{\partial (c_{2,v}A)}{\partial A},..., \frac{\partial (c_{i-j+1,v}A)}{\partial A}\right)\right) B_{n,i}(c_{1,b},c_{2,b},..., c_{n-i+1,b}).
\end{align}

\clearpage
\section{Additional results}\label{sec:app2}

\begin{figure}[h!] 
\begin{center}
\includegraphics[width=0.9\linewidth]{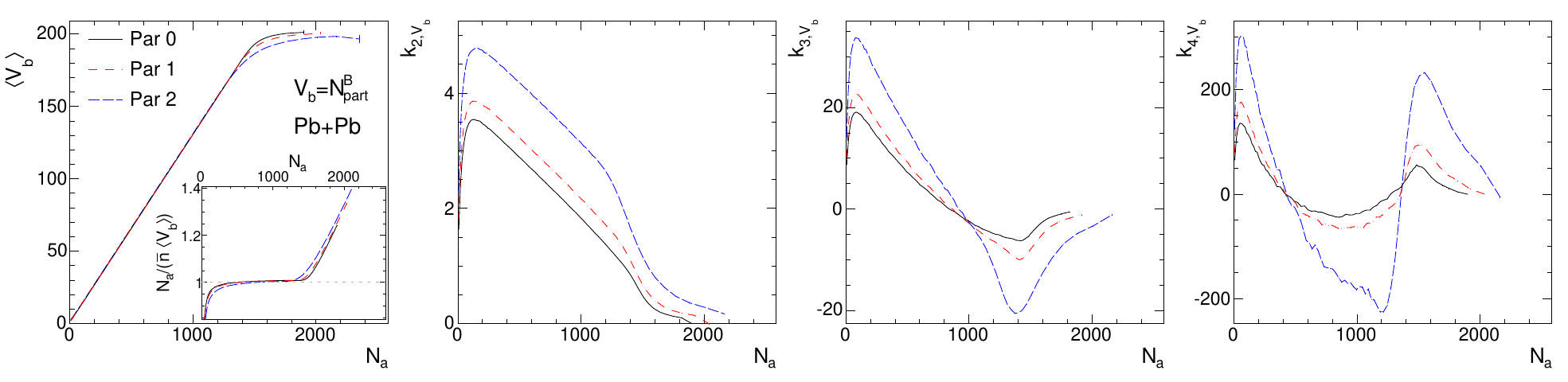}\\
\includegraphics[width=0.9\linewidth]{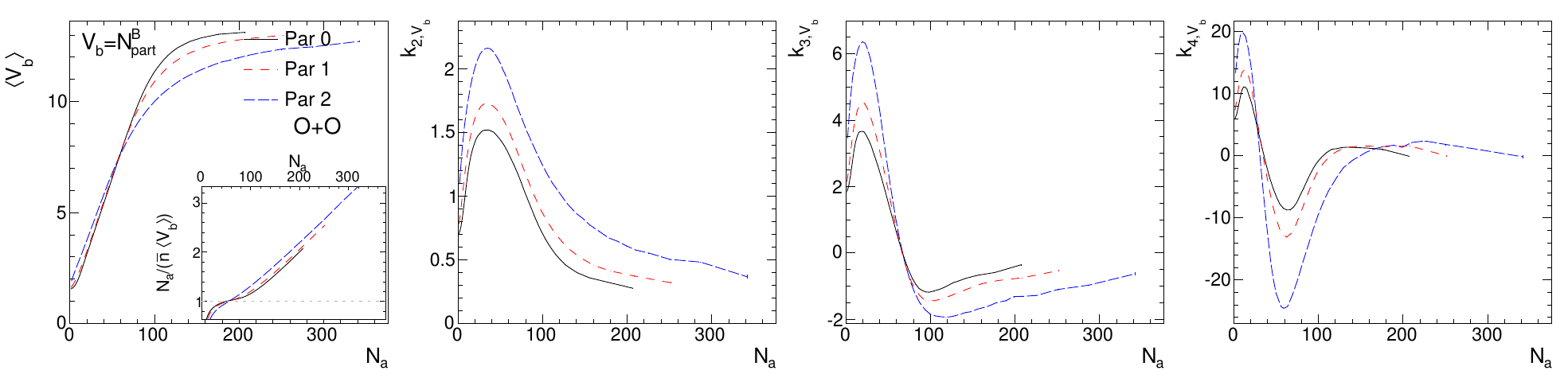}
\end{center}
\vspace*{-0.5cm}
\caption{\label{fig:app2} The cumulants $\kn{n}{\Nsb}$ as a function of $\Na$ to quantify the influence of centrality decorrelations in Pb+Pb system (top row) and O+O system (bottom row). They are calculated with $\Nsa=\npartf$ and $\Nsb=\npartb$ for the three parameter sets, with  $n=1$, 2, 3, and 4 from left to right panels. The insert panels show the ratio of $\Na/(\bar{n}\lr{V})$, to quantify the deviation from the linear relation.}
\end{figure}

\begin{figure}[h!]
\begin{center}
\includegraphics[width=0.9\linewidth]{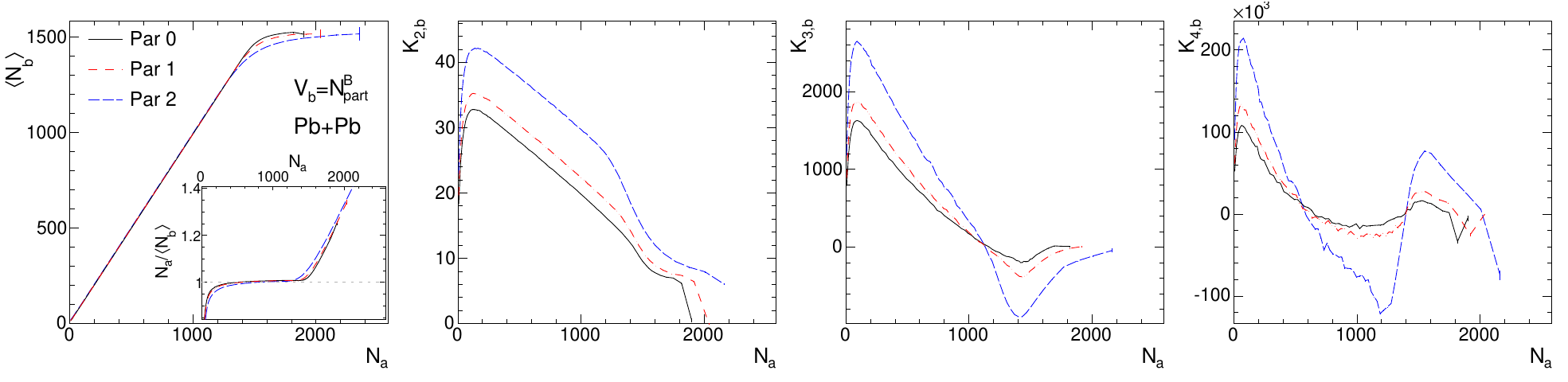}\\
\includegraphics[width=0.9\linewidth]{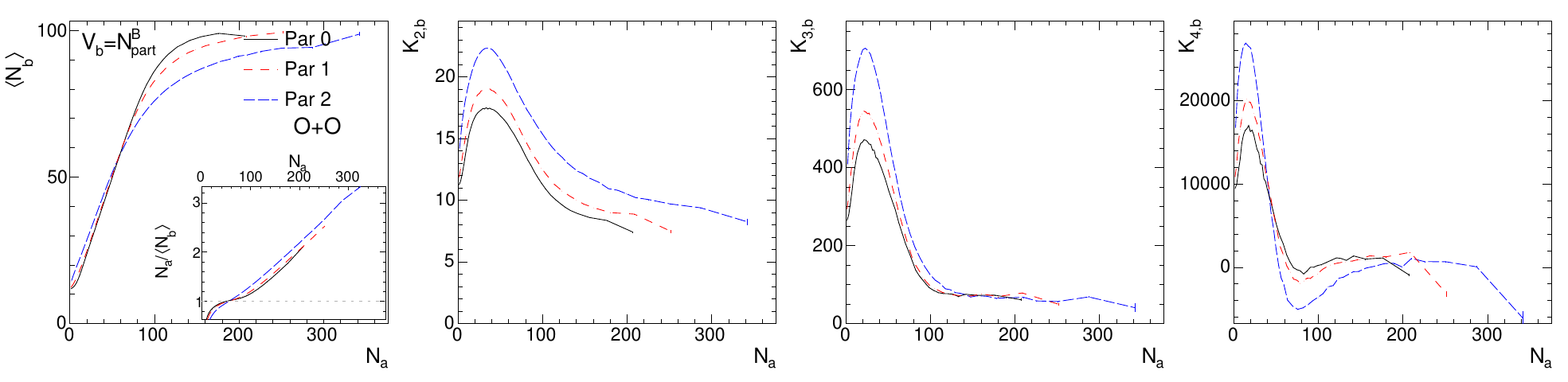}
\end{center}
\vspace*{-0.5cm}
\caption{\label{fig:app3} The cumulants of total multiplicity in subevent B $K_{n,b}$ as a function of $\Na$ to quantify the influence of centrality decorrelations in Pb+Pb system (top row) and O+O system (bottom row). They are calculated with $\Nsa=\npartf$ and $\Nsb=\npartb$ for the three parameter sets, with  $n=1$, 2, 3, and 4 from left to right panels. The insert panels show the ratio of $\Na/(\bar{n}\lr{V})$, to quantify the deviation from the linear relation.}
\end{figure}

Figures~\ref{fig:app2} and \ref{fig:app3} compare the $\kn{n}{\Nsb}$ and $K_{n,b}$, respectively, as a function of $\Na$ for the three parameter sets. This relation is described by Eq.~\eqref{eq:11}.

\begin{figure}[h!]
\begin{center}
\includegraphics[width=0.8\linewidth]{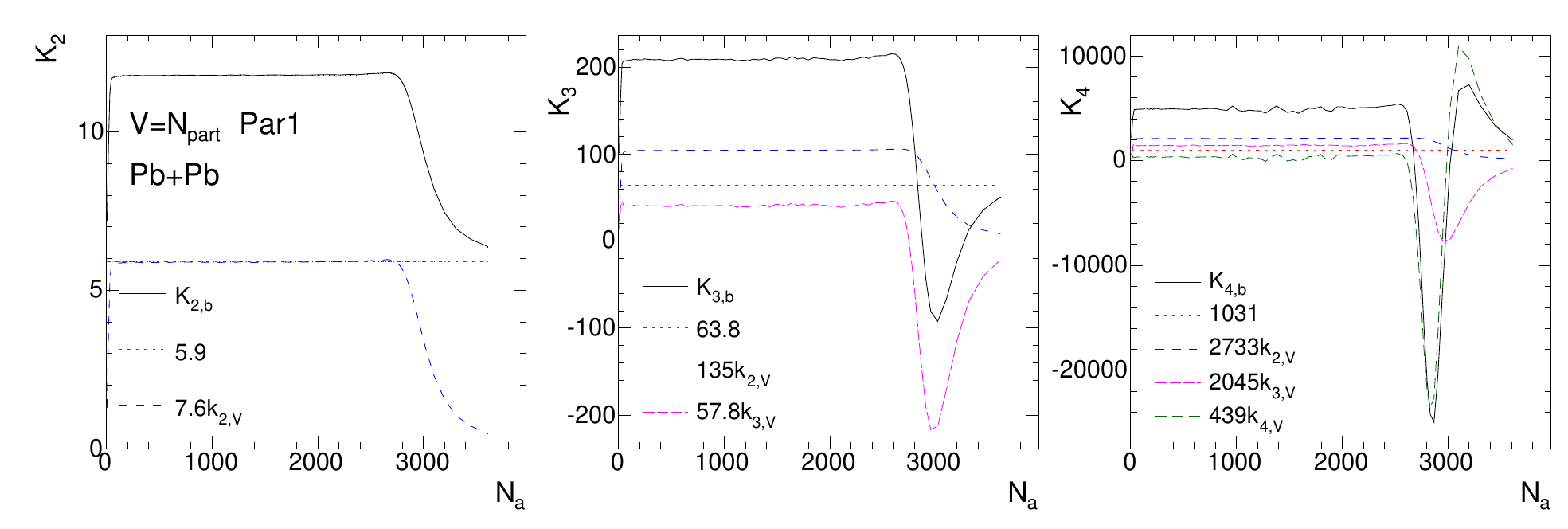}\\
\includegraphics[width=0.8\linewidth]{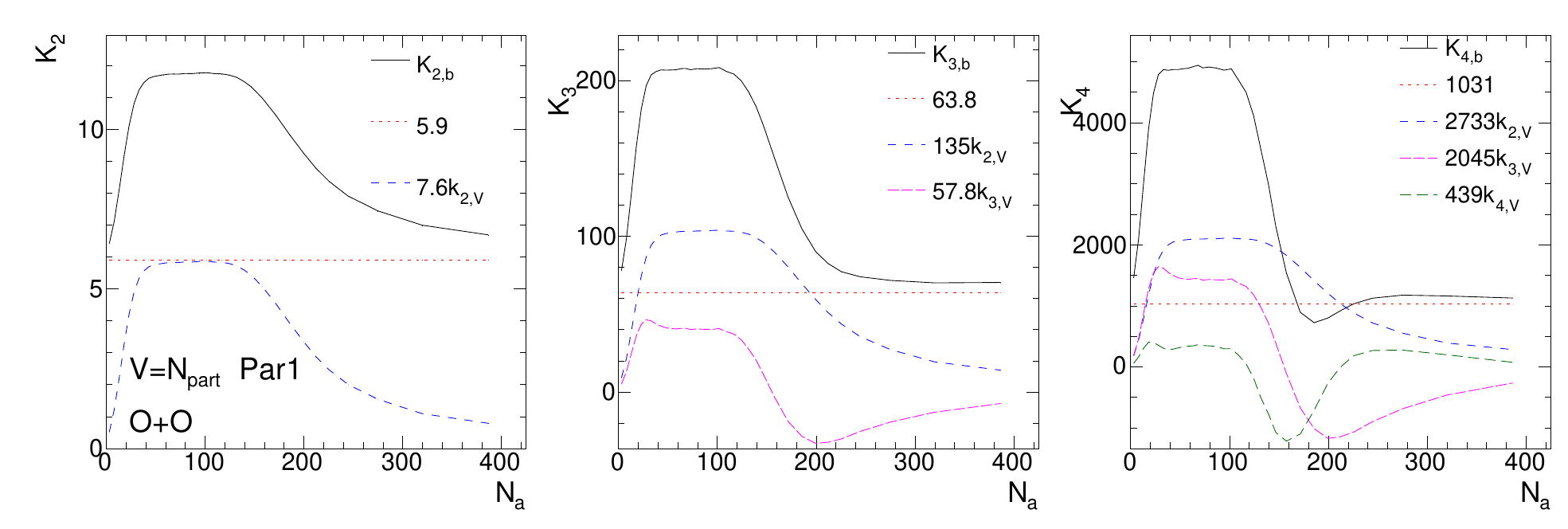}
\end{center}
\vspace*{-0.5cm}
\caption{\label{fig:app4} Decomposition of $K_{n,b}$ containing only multiplicity smearing effects as a function of $\Na$ into contributions from various terms according to Eq.~\eqref{eq:11}. The $K_{n,b}$ are obtained with $\Ns=\npart$ for Par1 in Pb+Pb (top row) and O+O systems (bottom row) with $n=2$, 3, and 4 from left to right panels.}
\end{figure}

\begin{figure}[h!]
\begin{center}
\includegraphics[width=0.8\linewidth]{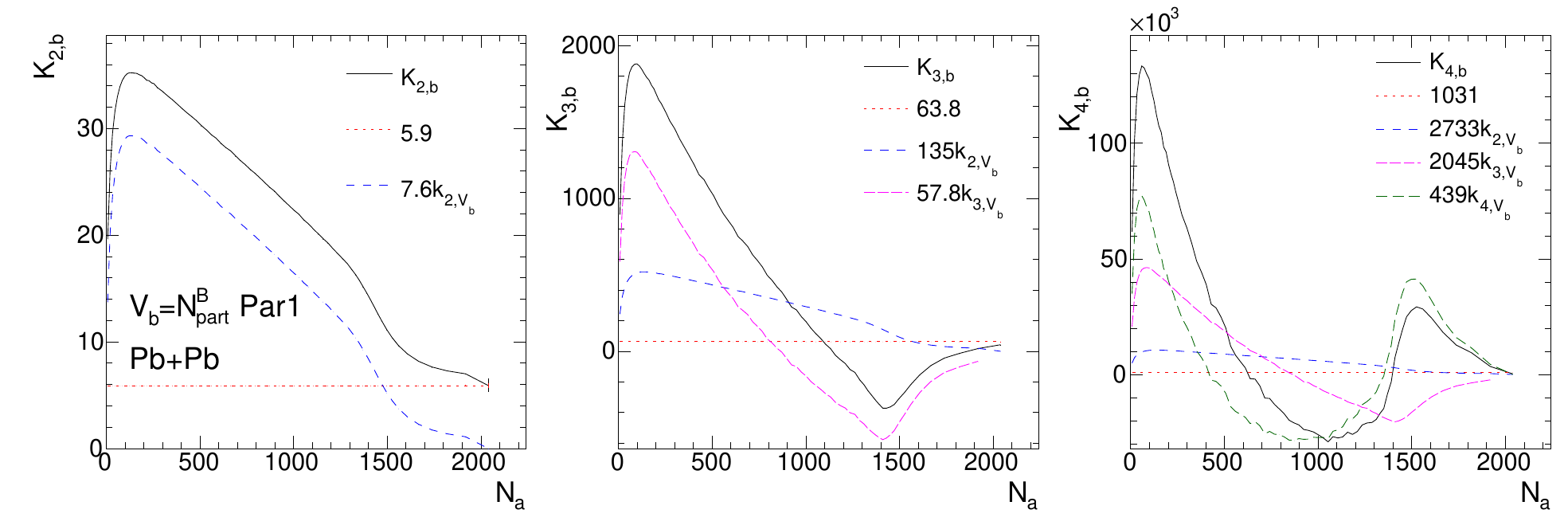}\\
\includegraphics[width=0.8\linewidth]{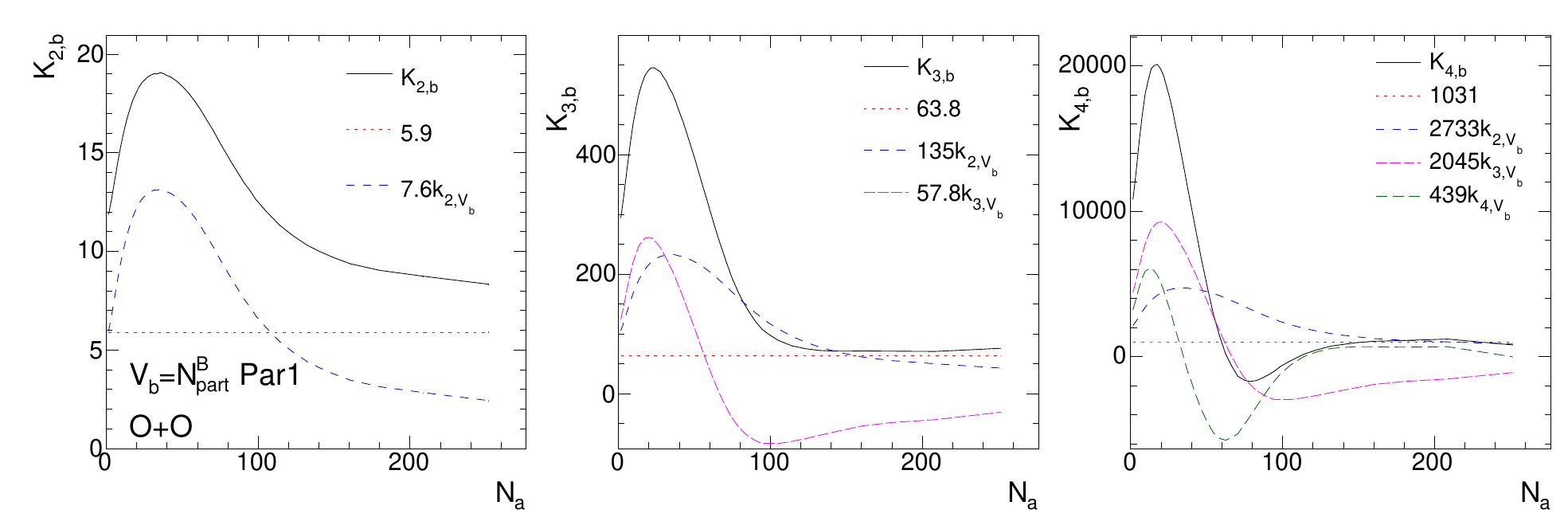}
\end{center}
\vspace*{-0.5cm}
\caption{\label{fig:app5} Decomposition of $K_{n,b}$ containing centrality decorrelation effects as a function of $\Na$ into contributions from various terms according to Eq.~\eqref{eq:11}. The $K_{n,b}$ are obtained with $\Nsa=\npartf$ and $\Nsb=\npartb$ for Par1 in Pb+Pb (top row) and O+O systems (bottom row) with $n=2$, 3, and 4 from left to right panels.}
\end{figure}

Figure~\ref{fig:app4} shows the decomposition of $K_{n,b}$ for the case of considering only multiplicity smearing and Figure \ref{fig:app5} shows the decomposition of $K_{n,b}$ considering also centrality decorrelations i.e $\Nsa=\npartf$ and $\Nsb=\npartb$. 

\clearpage
\bibliography{centdecor}{}
\bibliographystyle{apsrev4-1}

\end{document}